\documentclass[preprint,aip,jcp,superscriptaddress]{revtex4-1}

\usepackage{graphicx,color}
\usepackage{amsmath,amsfonts,amssymb}
\usepackage{mathtools}
\usepackage{dcolumn}
\usepackage{lmodern}
\usepackage{comment}
\usepackage[normalem]{ulem}
\usepackage{cancel}
\usepackage{xspace}
\usepackage{mhchem}

\newcommand{\fig}{Fig.}

\newcommand{\figref}[1]{\fig~\ref{#1}}

\newcommand{\tabref}[1]{table~\ref{#1}}

\renewcommand{\eqref}[1]{Equation~(\ref{#1})}

\newcommand{\ad}[1]{#1}

\newcommand{\rem}[1]{}
\newcommand{\stkout}[1]{\ifmmode\text{\sout{\ensuremath{#1}}}\else\sout{#1}\fi}

\newcommand{\chemshell}{ChemShell\xspace}
\newcommand{\dl}{DL-FIND\xspace}

\begin{document}
	
	\title{Gaussian Process Regression for Geometry Optimization}
	\author{Alexander Denzel}
	\affiliation{Institute for Theoretical Chemistry, University of Stuttgart, 
		Pfaffenwaldring 55, 70569 Stuttgart,Germany, kaestner@theochem.uni-stuttgart.de}
	
	\author{Johannes K\"{a}stner}
	\affiliation{Institute for Theoretical Chemistry, University of Stuttgart,
		Pfaffenwaldring 55, 70569 Stuttgart,Germany, kaestner@theochem.uni-stuttgart.de}
	
	
	\begin{abstract}
		We implemented a geometry optimizer
		based on Gaussian process regression (GPR) to find minimum structures
		on potential energy surfaces.
		We tested both a two times differentiable 
		form of the Mat\'{e}rn kernel and the squared exponential
		kernel. The Mat\'{e}rn kernel performs much better.
		We give a detailed description 
		of the optimization procedures. These include
                overshooting the step resulting from GPR in order
                to obtain a higher degree of interpolation
                vs. extrapolation. In a 
		benchmark against the 
		L-BFGS  
		optimizer of the \dl library on $26$ test systems,
		we found the new optimizer to generally reduce the
                number of required optimization steps. 
	\end{abstract}
	
	
	\maketitle

	\section{Introduction}
	Geometry optimization is one of the most 
	essential tasks in theoretical chemistry.
	Thereby, one has to reduce the amount of
	evaluations of the potential energy surface 
	(PES) as much as possible. 
	This is because energy evaluations can be quite costly, 
	depending on the chosen electronic structure method. 
	Nevertheless, most geometry optimizers are gradient-based:
	The gradient gives a good first estimate 
	of the direction in which 
	one can find, for example, a minimum.
	Most simply one can go in 
	the opposite direction of the gradient, 
	following the steepest descent.
	More frequently employed methods are 
	subspace methods, often using the Krylov subspace,
	like the conjugate gradient method.\cite{CG_Method}
	Second order optimizers, like Newton's method, converge much faster; 
	but the necessary generation of 
	Hessians is often too costly in quantum chemistry.
	Quasi-Newton methods use only gradient information,
	but try to build up an approximation of the inverse Hessian matrix 
	during the optimization procedure,
	which results in a considerable speedup. 
	Maybe the most popular variant of these methods is the 
	Limited-memory Broyden--Fletcher--Goldfarb--Shanno (L-BFGS) 
	algorithm.\cite{Broyden:1970:CCDa, fletcher1970new, goldfarb1970family, 
		shanno1970conditioning, fletcher1980practical, Liu1989,Nocedal1980}
	
	The main goal in the development of new geometry optimizers
	is to limit the number of energy/gradient 
	evaluations further and further.
	With increasing popularity of machine learning methods like 
	neural networks\cite{Behler2016}, and kernel methods\cite{Habershon_Tew,Bartok2013,ramakrishnan2017} in theoretical chemistry,
	the question arises whether these
	methods can be exploited to increase the efficiency of geometry optimizers. 
	For example, the nudged elastic band method
	\cite{MillsNEB_brief, HenkelmanNEB}
	was improved recently through the kernel-based methodology of 
	Gaussian process regression (GPR).\cite{GPRNEB_Jonsson}
	\ad{Furthermore, the method was used to fit multipole
		moments,\cite{IntramolMultipolKriging}
		polarizable water,\cite{polWaterKriging} predicting kinetic
		energies\cite{PredKinEnergyOfCoordsKriging}
		and many other chemical properties.
		\cite{Ramakrishnan2015,hansen2015interaction,PESandVibLevels}}
	In this paper we present a new gradient-based geometry optimizer
	that employs GPR to find minimum structures.
	\ad{For that we build a machine learned surrogate model
	for the PES that is improved on the fly. A similar approach was
        suggested previously
	to perform molecular dynamics calculations.\cite{OnTheFlyMachineLearningQMForces}}
	The resulting algorithm \ad{of our optimizer} 
	is implemented in the open-source
    optimization library, \dl\cite{dlfind}.
    Therefore, it may be used in
    \chemshell.\cite{SHERWOOD20031,ChemshellReview} It will be
    made available to the scientific community.

	This paper is organized as follows.
	We give a short introduction
	to the theory of GPR in Section~\ref{sec::theory_GPR}.
	In Section~\ref{sec::gpr_optimizer} we explain the implementation of
	our new optimizer,
	show the difficulties in the endeavor of using GPR for 
	geometry optimization, and how to overcome them.
	We also present benchmarks to compare our 
	GPR optimizer to the well established L-BFGS optimizer of 
	\dl in Section~\ref{sec::applications}.
	
	All properties in this paper are 
	expressed in atomic units (Bohr for positions and distances, Hartree for
        energies), unless other units are specified.

	\section{Theory}
	\label{sec::theory}
	GPR is a kernel-based machine learning technique 
	that uses the methodology of statistical/Bayesian inference.
	In this section we give a short introduction to this method.
	A set of so called \emph{training points} 
	at which we calculated the energy, and gradient 
	of the PES, is interpolated to infer the shape of the 
	PES in the vicinity of these points.
	We will first consider GPR by only using energy values.
	The inclusion of gradient information will be done in Section~\ref{sec::inludeDerivInfo}.
	Subsequently, we discuss the applicability for geometry optimization,
	and present the key elements of our GPR optimizer.
	
	\subsection{Gaussian process regression}
	\label{sec::theory_GPR}	
	To clarify what a Gaussian process is, 
	we will first define
	the more general term, \emph{stochastic process}:
	A stochastic process is a collection of random variables. 
	If we want to represent a PES
	as a stochastic process, 
	we can assign every point, i.e. every molecular geometry,
	in the configuration space a random variable. 
	This random variable will take on the value of an
	energy.
	Giving the joint probability distribution for every 
	finite subset of these random variables, specifies 
	the stochastic process.
	If these distributions are multivariate Gaussian distributions,
	the stochastic process is called a Gaussian process (GP). 
	Given the fixed energy values at some training points,
	this yields a probability distribution
	over the energy value at
	any point in configuration space.
	From that distribution we can estimate 
	the most probable energy value at this point,
	and also how likely this value is.
	One can use arbitrary coordinate systems for GPR, but the
	coordinate system will influence the resulting parameters
	and accuracy.

	Initially we restrict ourselves to the task of 
	interpolating a PES from some given energy values.
	A GP is uniquely defined by a so called \emph{prior mean function},
	and the \emph{covariance function}.
	The prior mean function is a guess of the PES before 
	one has included any training points in the scheme. 
	It can be a sophisticated estimate of the PES, but often 
	one simply uses zero, or the constant average 
	value of all energy values in the training set 
	as the prior mean. 
	The covariance function, $k(\vec{x},\vec{x}^{\, \prime})$,
	describes the covariance between the two random variables 
	specified by the coordinates, $\vec{x},\vec{x}^{\, \prime}\in \mathbb{R}^{d}$, of a $d$ dimensional system.
	These can be the $d=3n$ Cartesian coordinates of 
	the $n$ atoms in a molecule. 
	We also call the covariance function \emph{kernel}, although not every
	kernel has to be a covariance function:
	It is not mandatory for a kernel to
	describe a covariance of random variables.
	But in the framework of GPR 
	one wants to keep the interpretation of 
	the covariance; therefore, one must  
	choose a covariance function as a kernel.
	A kernel is a valid covariance function, 
	if and only if it is symmetric, 
	i.e. $k(\vec{x},\vec{x}^{\, \prime})=k(\vec{x}^{\, \prime},\vec{x})$, 
	and
	\begin{align}
	\sum_{i=1}^{N}\sum_{j=1}^{N}c_i k(\vec{x}_i,\vec{x}_j)c_j
	\end{align}
	is non-negative for all $N\in\mathbb{N}$, 
	all $\vec{x}_i,\vec{x}_j\in \mathbb{R}^{d}$,
	and all coefficients $c_k \in \mathbb{R}$ for
	$k=1, ..., N$.
	
	Here we consider only two specific covariance functions.
	These are stationary isotropic, 
	i.e. they only depend on $r=|\vec{x}-\vec{x}^{\, \prime}|$
	which we will interpret as the Euclidean distance 
	between the two points $\vec{x}$ and $\vec{x}^{\, \prime}$.
	Furthermore, the closer/distant a training point is
	the larger/smaller its influence on the 
	estimate of the energy
	should be. This influence is represented
	by the covariance function.
	Consequently, the covariance function 
	should decrease with increasing 
	distance, $r$.
	
\begin{figure}
		\begin{center}
			\includegraphics[width=8cm]{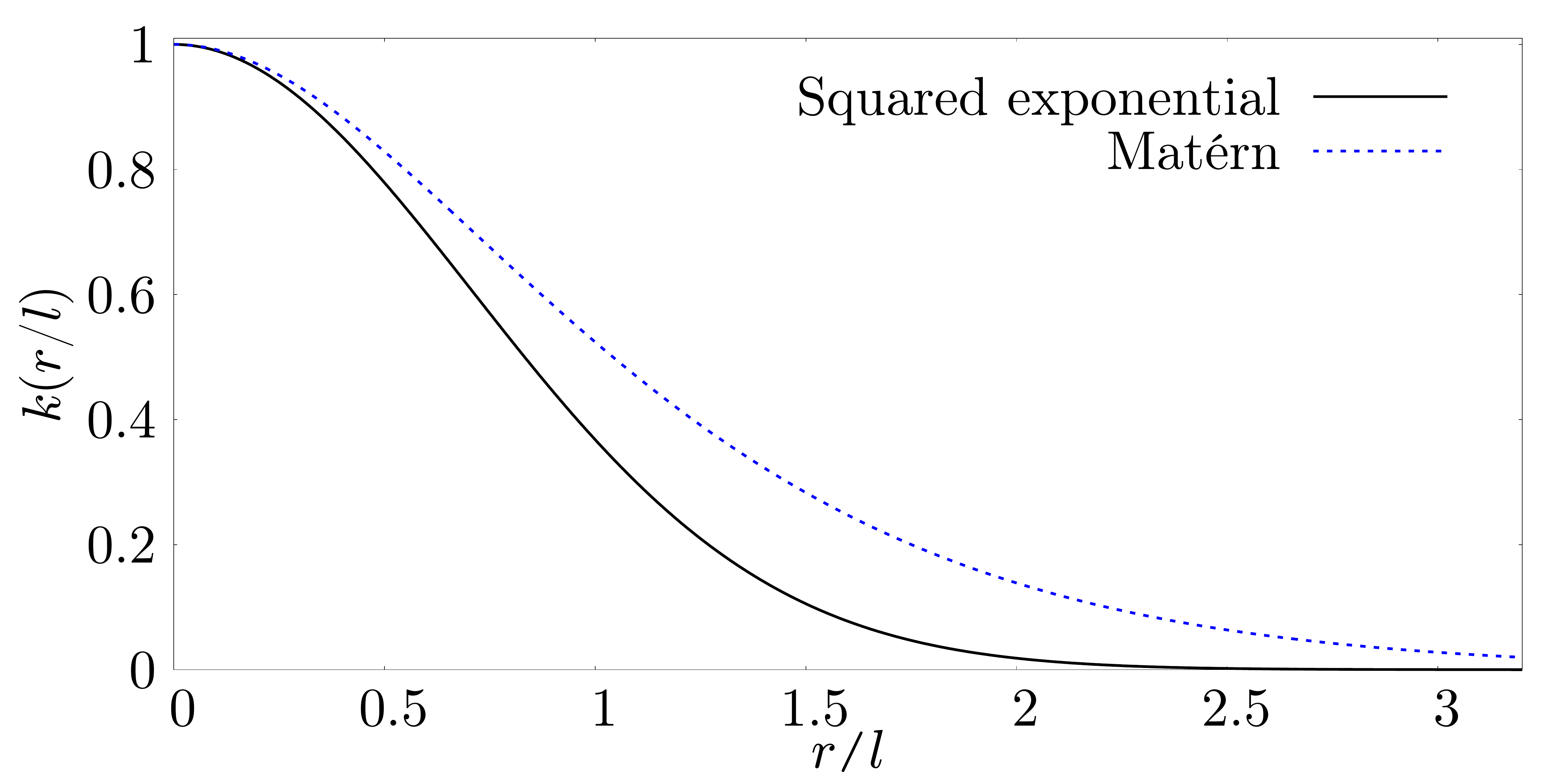}
			\caption{The squared exponential covariance 
				function from \eqref{eq::SE_Kernel}
				and the Mat\'{e}rn covariance function from 
				\eqref{eq::M_Kernel} with parameters
				$l,\sigma_\text{f}=1$.
			}
			\label{fig::compareSEandM}
		\end{center}
	\end{figure}

	We consider the squared exponential covariance function
	\begin{equation}
	\label{eq::SE_Kernel}
	k_{\text{SE}}(r)=
	\sigma_\text{f}^2\exp\left(-\frac{r^2}{2l^2}\right)
	\end{equation}
	and a form of the Mat\'{e}rn covariance function\cite{matern2013spatial}
	\begin{equation}
	\label{eq::M_Kernel}
	k_{\text{M}}(r)=\sigma_\text{f}^2\left(1+\frac{\sqrt{5}r}{l}+\frac{5r^2}{3l^2}\right)\exp\left[-\frac{\sqrt{5}r}{l}\right]
	\end{equation}
	both of which we implemented.
	The functions are depicted in \figref{fig::compareSEandM}.
	The parameter $\sigma_\text{f}$
	could be used to maintain numerical stability by 
	scaling up the value of covariances, but it will have no
	influence on the analytical solution.
	We simply chose it to be $1$ in our algorithm.
	The parameter $l$ defines a certain
	characteristic length-scale of the GP. 
	It will have the biggest influence on the
	obtained interpolant since it 
	defines the sphere of influence
	that a training point will have in the GP.
	The presented two covariance functions are especially 
	interesting because they guarantee a PES
	that is in the $C^2$ class, 
	i.e. two times continuously differentiable.\cite{rasmussen2006gaussian}
	The squared exponential covariance function implies that the PES
	is also in the $C^\infty$ class. We usually assume this to be true
	in theoretical chemistry, but our tests show that
	the presented Mat\'{e}rn covariance function 
	yields better results.
	This is because the high constraint that
	the smoothness of $k_{\text{SE}}$ implies on the
	GP leads to overshooting and 
	oscillation, especially in the close
	extrapolation regime.
	Additionally, in our experiments we found that one should
	choose a larger $l$ for the 
	Mat\'{e}rn kernel than for the squared exponential kernel.

	To model a PES with a GP we take a training set including $N$ 
	configurations of the molecule $\vec{x}_1,\vec{x}_2,...,\vec{x}_N$, and the respective energies at these points
	$E_1,E_2,...,E_N$, so that the calculated electronic structure
	yields the energy $E_m$ given the configuration $\vec{x}_m$ of the atoms.
	Note that the points $E_m$ are the energy values within the
	used electronic structure method.
	Consequently, the energies may be noisy.
	In GPR one assumes a normally distributed noise 
	on the values $E_m$ with 
	its variance given by a parameter $\sigma_\text{e}^2$.
	
	We also introduce the prior mean function
	$E_{\text{prior}}(\vec{x})$ which is our estimate of the PES
	before we have included any training points.
	Consequently, 
	GPR only learns the error of $E_{\text{prior}}(\vec{x})$ rather
	than the PES directly.
	The prior mean function can also be considered 
	as a bias that mostly affects the
	regression scheme away from the training points.
		
	A GP yields a \emph{posterior mean} function $E(\vec{x})$, 
	which is the most probable
	value of the energy at the position $\vec{x}$ 
	in the stochastic GP framework;
	therefore, also the prediction of the GPR-PES.
	It is given by \cite{rasmussen2006gaussian}
	\begin{equation}
	\label{eq::GPprediction}
	E(\vec{x})= \sum_{n=1}^{N}w_n k(\vec{x},\vec{x}_n) + E_{\text{prior}}(\vec{x})
	\end{equation}
	in which $\vec{w}=(w_1\ w_2\ ...\ w_N)^T$ is the solution of the linear system
	\begin{equation}
	\label{eq::linSystem}
	\sum_{n=1}^{N}K_{mn}w_n=E_m - E_{\text{prior}}(\vec{x}_m)
	\end{equation}
	for all $m=1,2,...,N$ and 
	\begin{equation}
	\label{eq::cov_mat_elements}
	K_{mn}=k(\vec{x}_m,\vec{x}_n)+\sigma_\text{e}^2\delta_{mn}
	\end{equation}
	are the entries of the so called \emph{covariance matrix} $K$ which 
	contains the covariances between the training data and $\delta_{mn}$
	which is the Kronecker delta.
	The parameter $\sigma_\text{e}$ arises from the already mentioned,
	normally distributed noise in the $E_m$ with variance $\sigma_\text{e}^2$.
	If we use a kernel that is a covariance function,
	the covariance matrix $K$ is positive semi-definite,
	which follows from the properties of the covariance
	as a positive semi-definite symmetric bilinear form.
	Note that the computationally most demanding step
	is to solve the linear system of \eqref{eq::linSystem}.
	In our code this is carried out via a standard
	Cholesky decomposition.
	The computational effort to solve such a linear
	system scales cubically 
	with the number of training points.
	
	\subsection{Including derivative information\label{sec::inludeDerivInfo}}
	Derivation is a linear operation.
	Therefore, the derivative of a GP is again 
	a GP.\cite{rasmussen2006gaussian}
	Random variables in the stochastic process can,
	therefore, also take on the values of a derivative
	of the energy instead of the energy itself.
	Let us consider a Gaussian process with an arbitrary
	kernel function, $k(\vec{x},\vec{x}_n)$.
	The interpolant (in our case for example
	the energy, $E$) is calculated 
	according to \eqref{eq::GPprediction}.
	Since the kernel function is the only dependency on $\vec{x}$
	we can use learned derivative information
	in GPR. For example, the expression
	\begin{equation}
	\tilde{E}^i(\vec{x})= \sum_{n=1}^{N}v_n^i \frac{dk(\vec{x},\vec{x}_n)}{dx_n^i}
	\end{equation}
	is also the prediction of 
	a Gaussian process with a different covariance function
	$\frac{dk(\vec{x},\vec{x}_n)}{dx_n^i}$,
	that exploits information about the derivative at the training point 
	$\vec{x}_n$ for inference. We use the notation
	$\frac{d}{dx_n^i}$ as the derivative with respect to
	the variable $\vec{x}_n$
	in the direction of the unit vector in the $i$-th dimension.
	Combining learned energies and gradients,
	the inferred GPR-PES is
	\begin{equation}
	\label{eq:GPREnergy_with_Derivatives}
	E(\vec{x}) = \sum_{n=1}^{N} w_n k(\vec{x}, \vec{x}_n) + 
	\sum_{n=1}^{N} \sum_{i=1}^{d}
	v_n^i \frac{dk(\vec{x},\vec{x}_n)}{dx_n^i} + E_{\text{prior}}(\vec{x})
	\end{equation}
	with new parameters $v_n^i$.
	Our GPR optimizer is based on this equation.
	The $k$-th element of the gradient on this
	GPR-PES can also be obtained analytically.
	\begin{equation}
	\begin{split}
	\frac{d}{dx^k} E(\vec{x}) = &\sum_{n=1}^{N} w_n 
	\frac{d}{dx^k}k(\vec{x}, \vec{x}_n) + 
	\sum_{n=1}^{N} \sum_{i=1}^{d}
	v_n^i \frac{d^2k(\vec{x},\vec{x}_n)}{dx^kdx_n^i} \\
	&+\frac{d}{dx^k} E_{\text{prior}}(\vec{x})
	\end{split}
	\end{equation}
	The weights $w_n$ and $v_n^i$ will be obtained 
	by solving a larger linear system with 
	a covariance matrix of the following form.
	\begin{align}
	\label{eq:KM_firstDeriv}
	K=\begin{bmatrix}
	k(\vec{x}_m,\vec{x}_n)+\sigma_\text{e}^2\delta_{mn} & \frac{dk(\vec{x}_m, \vec{x}_n)}{dx_n^i}\\
	\frac{dk(\vec{x}_m, \vec{x}_n)}{dx_m^i} & 
	\frac{d^2k(\vec{x}_m, \vec{x}_n)}{dx_m^idx_n^j}+\sigma_\text{g}^2 \delta_{mn}\delta_{ij}
	\end{bmatrix}
	\end{align}
	The parameter $\sigma_\text{g}$ describes the variance
	of the assumed normally distributed noise on the 
	input gradient entries. This is 
	equivalent to $\sigma_\text{e}$ 
	for the energy values.
	The linear system of \eqref{eq::linSystem}
	will become
	\begin{equation}
	\label{eq::linSystemWithGradients}
	K\begin{bmatrix}
	w_1 \\
	\vdots\\
	w_N \\
	\vec{v}_1\\
	\vdots\\
	\vec{v}_N
	\end{bmatrix}=
	\begin{bmatrix}
	E_1 \\
	\vdots\\
	E_N \\
	\vec{g}_1\\
	\vdots\\
	\vec{g}_N
	\end{bmatrix}
	-
	\begin{bmatrix}
	E_{\text{prior}}(\vec{x}_1) \\
	\vdots\\
	E_{\text{prior}}(\vec{x}_N) \\
	\vec{\nabla} E_{\text{prior}}(\vec{x})|_{\vec{x}=\vec{x}_1}\\
	\vdots\\
	\vec{\nabla} E_{\text{prior}}(\vec{x})|_{\vec{x}=\vec{x}_N}
	\end{bmatrix}
	\end{equation}
	in which $E_m$ is the energy and $\vec{g}_m$ 
	is the gradient at point $\vec{x}_m$, and
	$\vec{\nabla} E_{\text{prior}}(\vec{x})|_{\vec{x}=\vec{x}_m}$
	is the gradient of the prior mean function at the point
	$\vec{x}_m$,
	and $\vec{v}_m=\begin{bmatrix}
	v_m^1 & v_m^2 & \dots & v_m^N
	\end{bmatrix}^T$ contains the coefficients from above.
	Note that the covariance matrix in
	this linear system
	has the size $N(d+1) \times N(d+1)$.
	Therefore, the required CPU time to solve
	\eqref{eq::linSystemWithGradients}
	formally scales
	with $\mathcal{O}(\left[N(d+1)\right]^3)$,
	if we solve the linear system exactly.
	To overcome this obstacle we will introduce
	an approach that uses multiple GP layers 
	to bring the scaling down, 
	see Section~\ref{sec::multiLevelGPR}.
	Iterative solution of the linear system,
	does not easily give a good enough solution
	since the covariance matrix is not necessarily
	diagonally dominant.
	To see how to use GPR with second
	order derivatives we provide additional
	information in the supplementary material.
	
	\subsection{The GPR optimization}
	\label{sec::gpr_optimizer}
	The basic idea of our GPR optimizer is to use already obtained
	energy and gradient information of the PES
	to build a GP surrogate for it, the GPR-PES.
	Then we search for a minimum on this GPR-PES
	to estimate a minimum on the real PES.
	This process is repeated until we consider the optimizer 
	to be converged.
	So far, this is similar to other optimizers
	with different surrogate models
	based on Taylor expansions instead of
	GPR.\cite{ZhengOptInterpolSurfaces,Shepard}
	We first explain the optimization procedure in detail.
	In Section~\ref{sec::AlgorithmInOneD}
	we show an example for
	the resulting algorithm in one dimension.
	
	In order to define convergence for our optimizer
	we use the standard convergence
	criteria of \dl for the step size and the gradient:
	The Euclidean norm of the step vector, and the gradient vector,
	as well as the maximum entry of both vectors have
	to drop below a certain threshold.
	The step vector is simply the vector
	that points from the last estimate to the
	new estimate of the minimum, and
	describes the proceeding of the optimization run.	
	Given a single tolerance value, $\delta$, the
	convergence criteria in \dl are
	\begin{alignat}{3}
	\label{eq::convCriteria1}
	\max\limits_i (g_i) &< \delta_{\mathrm{max}(g)} &&\coloneqq \delta \\
	\label{eq::convCriteria2}
	\frac{|\vec{g}|}{d} &< \delta_{|g|} &&\coloneqq \frac{2}{3}&&\delta \\
	\label{eq::convCriteria3}
	\max\limits_i (s_i) &< \delta_{\mathrm{max}(s)} &&\coloneqq 4&&\delta \\
	\label{eq::convCriteria4}
	\frac{|\vec{s}|}{d} &< \delta_{|s|} &&\coloneqq \frac{8}{3}&&\delta
	\end{alignat}	
	where $|\vec{g}|$ ($|\vec{s}|$) is the Euclidean norm of
	the gradient (step vector), and $\max_i (g_i)$ 
	($\max_i (s_i)$) its
	maximum entry.
	If these four criteria are fulfilled,
	the algorithm is considered to be converged. Note that
        convergence is tested for the gradient on the underlying
        ab-initio data rather than the GPR fit.
		
\begin{figure}[h!]
		\begin{center}
			\includegraphics[width=8cm]{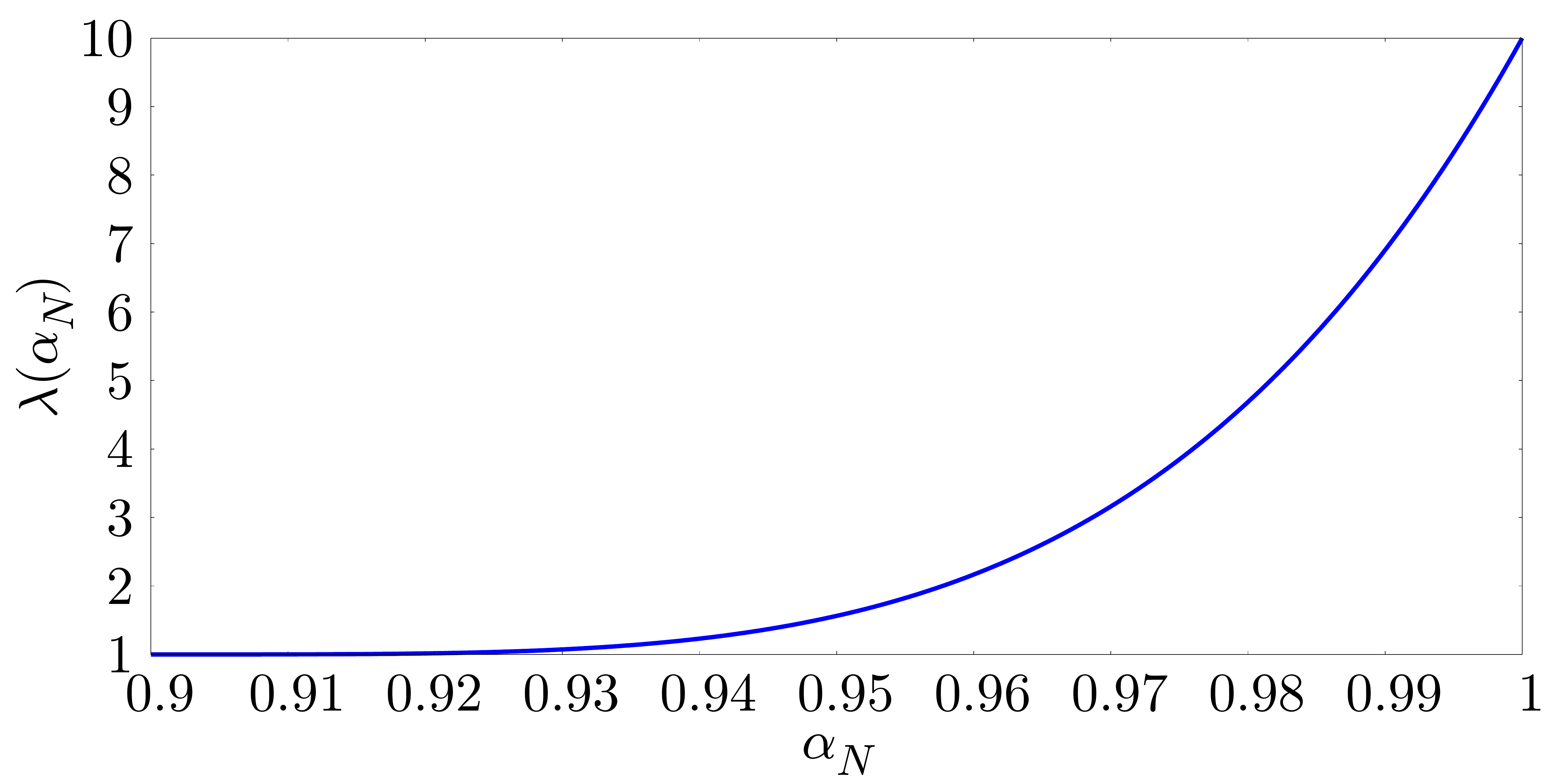}
			\caption{The scaling factor for overshooting the step, 
				see \eqref{eq::overshooting},
				with $\lambda_{\mathrm{max}}=10$ is plotted
				against $\alpha_N$.}
			\label{fig::MaxScalingLimitation}
		\end{center}
	\end{figure}

	In the first step the GPR-PES is built 
	by the energy, and the gradient 
	at one single starting point $\vec{x}_0$.
	In later steps we use all the obtained energies, and gradients
	to build the GPR-PES.
	We then find the minimum $\vec{x}^{\, \text{GPmin}}_N$
	on our GPR-PES with the already obtained $N$ training points.
	The GPR-PES is very cheap
	to evaluate, especially compared to the evaluation of the PES
	via electronic structure calculations. 
	Therefore, the search for $\vec{x}^{\, \text{GPmin}}_N$ can be carried out
	very fast with an arbitrary optimization method.
	In our case we use a L-BFGS optimizer\cite{Liu1989} for that matter.
	We usually start the search for a minimum on the GPR-PES
	at the last training point. 
	If the direction along the optimization is changed by
	more than a $90$ degree angle,
	or if the absolute value of the gradient gets larger,
	we search for a minimum, starting at each of the 
	$10\%$ of training points with lowest energies.
	The lowest minimum found is the next
	$\vec{x}^{\, \text{GPmin}}_N$.
	The obvious optimization step $\vec{s}^{\,\prime}_N$ 
	after one has obtained $N\geq 1$ training points
	would be to take the step vector to position 
	$\vec{x}^{\, \text{GPmin}}_N$ as the
	next guess for our minimum on the PES.
	\begin{equation}
	\label{eq::Step_noOvershoot}
	\vec{s}^{\, \prime}_N = \vec{x}^{\, \text{GPmin}}_N - \vec{x}_{N-1}
	\end{equation}
	The first training point $\vec{x}_{0}$ 
	is defined as the starting point of the 
	optimization, and $\vec{x}_{N-1}$ is then 
	the last estimate of the PES minimum when
	$N-1$ additional
	training points were obtained.
	This yields already a functional optimizer,
	but its performance is rather poor.
	This arises from a well known problem:
	GPR and other machine learning techniques
	are highly capable in interpolation, but 
	often perform quite poorly 
	in extrapolation. 
	An iterative optimization as described above is obviously
	largely based on extrapolation. 
	To represent the problem in a way that we mostly interpolate,
	rather than to extrapolate, we overshoot the estimated minimum on purpose:
	The first optimization step is carried out as described by 
	\eqref{eq::Step_noOvershoot}, and we define the first step as
	$\vec{s}_1=\vec{s}^{\, \prime}_1$.
	From the second step onward, however, we 
	determine the cosine of the angle between the last optimization step $\vec{s}_{N-1}$
	and the estimated new step $\vec{s}^{\, \prime}_N$
	\begin{equation}
	\label{eq::angle}
	\alpha_N = \frac{(\vec{s}_{N-1},\vec{s}^{\, \prime}_N)}{|\vec{s}_{N-1}||\vec{s}^{\, \prime}_N|}
	\end{equation}
	with $(\cdot,\cdot)$ being the Euclidean dot product.
	The closer $\alpha_N$ is to $1$, the smaller the angle
	becomes. If $\alpha_N$ is smaller than
	$0$, the direction of the optimization is changed
	by more than a $90$ degree angle. If it is close to $-1$,
	the direction is completely inverted.
	As soon as 
	\begin{equation}
	\label{eq::angle_startOvershooting}
	\alpha_N > 0.9
	\end{equation} 
	we scale up the initially estimated
	$\vec{s}^{\, \prime}_N$ to obtain the
	next optimization step
	\begin{equation}
	\vec{s}_N={\lambda}(\alpha_N)\vec{s}^{\, \prime}_N
	\end{equation}
	for $N\geq 2$ and introduce the scaling factor
	\begin{equation}
	\label{eq::overshooting}
	\lambda(\alpha_N) = 1+ ({\lambda}_{\mathrm{max}}-1)\left(\frac{\alpha_N-0.9}{1-0.9}\right)^4
	\end{equation}
	with a maximum scaling factor of ${\lambda}_{\mathrm{max}}$ so that
	$1\leq \lambda(\alpha_N)\leq {\lambda}_{\mathrm{max}}$.
	This scaling factor is depicted in \figref{fig::MaxScalingLimitation}.
	
\begin{figure}[h!]
			\begin{center}
				\includegraphics[width=8cm]{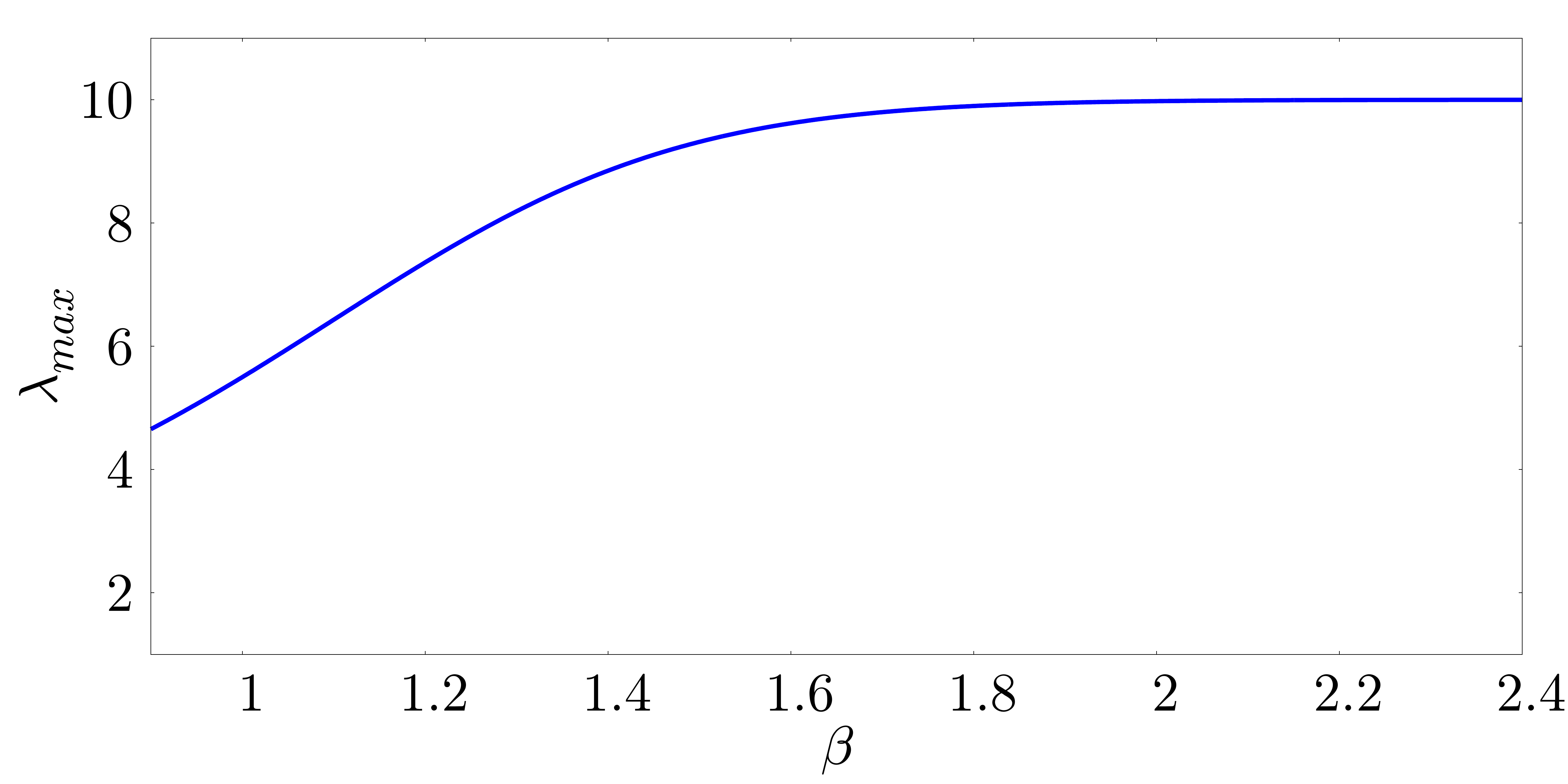}
				\caption{The limitation of the scaling factor, see
					\eqref{eq::MaxScalingLimitation}, is plotted
					against the variable $\beta$, see 
					\eqref{eq::MaxScalingLimitationBeta}.
					The upper limit of $\lambda_{\mathrm{max}}$ is set to
					$\tilde{\lambda}_{\mathrm{max}}=10$.
					}
				\label{fig::MaxScalingLimitationBeta}
			\end{center}
		\end{figure}

	To avoid large overshooting in the area around the 
	actual minimum on the PES we only apply this scaling
	factor, if the estimated step $\vec{s}^{\, \prime}_N$ does
	not satisfy the convergence criteria 
	of the maximum step entry in 
	\eqref{eq::convCriteria3}.
	Close to convergence we also limit the maximum scaling factor through
	\begin{equation}
	\label{eq::MaxScalingLimitation}
	\lambda_{\mathrm{max}} =\left(1+\tanh\left(\beta^{2}-1\right)\right)\frac{\tilde{\lambda}_{\mathrm{max}}-1}{2}+1
	\end{equation}
	in which 
	\begin{equation}
	\label{eq::MaxScalingLimitationBeta}
	\beta=\max\limits_{i} (s^{\prime}_i)/\delta_{\max (s)}
	\end{equation}
	is the ratio of the maximum entry of $\vec{s}^{\, \prime}$ and
	$\delta_{\max (s)}$,
	the convergence criterion for the maximum step entry
	from \eqref{eq::convCriteria3}.
	The variable
	$\beta$ indicates how close the algorithm is to
	convergence with respect to the 
	maximum entry of the step vector.
	If $\beta\leq 1$, the convergence criterion is met.
	No scaling will occur in his region.
	We plot $\lambda_{\mathrm{max}}$
	against $\beta$
	in \figref{fig::MaxScalingLimitationBeta}.
	
	To some extent, the limitation of 
	the highest possible scaling factor $\tilde{\lambda}_{\mathrm{max}}$
	to $\lambda_{\mathrm{max}}$
	via \eqref{eq::MaxScalingLimitation}
	is intended to guarantee a smooth transition into the 
	region of convergence. However, keeping at least
	$\tilde{\lambda}_{\mathrm{max}}/2$ at the point
	where the convergence criterion is met
	does not seem to hinder convergence.
	Consequently, we keep 
	$\lambda_{\mathrm{max}}\approx \tilde{\lambda}_{\mathrm{max}}/2$ in the
	area of convergence.
	The value of $\tilde{\lambda}_{\mathrm{max}}$ is chosen to be
	$5$ at the beginning of the optimization.
	It is increased by $5\%$, 
	if we observe that we do
	more than one overshooting according to \eqref{eq::overshooting}
	in a row, i.e. that the criteria of 
	\eqref{eq::angle_startOvershooting} are satisfied for two
	consecutive steps in the optimization procedure.
	At the end of each optimization step,
	the estimated step size $|\vec{s}_N|$ is limited
	by the maximum step size $s_{\mathrm{max}}$ 
	we set externally for the optimization procedure.\\

	\subsubsection{Separate dimension overshooting}
	\label{sec::sepDimOvershooting}
	In several tests we observed that usually only a few
	dimensions seem to converge very slowly, while
	the convergence in the other dimensions is already accomplished.
	This is especially the case in longer optimization runs.
	We assume that the reason for this may be that we 
	use only one length scale parameter $l$ in 
	\eqref{eq::M_Kernel}, and do not assume different length scales
	for different dimensions.
	On the other hand, it is not easy to find 
	suitable parameters for every dimension,
	and introducing more parameters lets the optimizer
	become more prone to chance.
	Instead we make use of the fact that we can overshoot the correct
	solution to our optimization problem quite a bit,
	and introduce an additional \emph{separate dimension overshooting}:
	We consider every dimension independently of the others.
	If we observe that the optimizer has monotonically changed
	the value of the coordinate in this dimension over the last $20$ 
	optimization steps, we build up a one dimensional
	GP to represent the optimization along this single coordinate.
	It approximates the value of the corresponding coordinate
	with respect to the number of steps taken:
	The position of the training points for this GP is simply
	the number of the step along the optimization procedure.
	The value that is interpolated 
	is the value of the considered coordinate 
	at that step.
	On this GP we search for the next maximum/minimum
	assuming it could be a good guess for the dimension's
	value at the real minimum of the PES.
	Thereby, we ignore coupling of the different
	coordinates. 
	To give the optimization procedure time to explore
	the omitted coupling, we suspend the 
	separate dimension overshooting for $20$ optimization steps
	after we performed it.
	To restrict the overshooting to a reasonable regime, we
	limit the separate dimension overshooting by a 
	factor of $4$ compared to the originally estimated
	step without any overshooting.
	Furthermore, we only apply the separate dimension overshooting,
	if it suggests higher overshooting than the scaling factor
	in \eqref{eq::overshooting}, and if the
	convergence criterion for the maximum step entry from 
	\eqref{eq::convCriteria3} is not satisfied.
	Also this overshooting procedure is finally limited
	by the maximum step $s_{\mathrm{max}}$ 
	allowed for the optimization procedure.

	\subsubsection{The algorithm in one dimension}
	\label{sec::AlgorithmInOneD}
\begin{figure}[h!]
		\begin{center}
		\includegraphics[width=8cm]{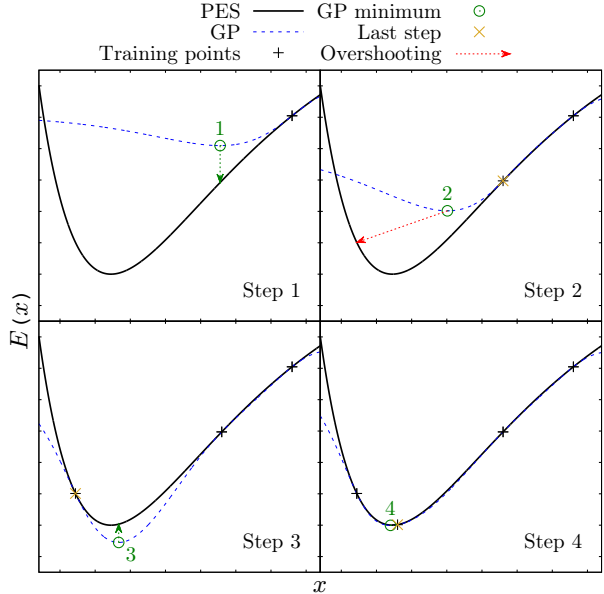}
		\caption{The basic idea of the GPR optimizer in the 
			case of a Lennard--Jones potential
			as a simple example for a PES.\\
		}
		\label{fig::GPR_opt_principle}
		\end{center}
	\end{figure}

	We explain the overall optimization 
	process in a simple one-dimensional 
	example PES $E(x)$ illustrated in \figref{fig::GPR_opt_principle}.
	\begin{itemize}
	\item \emph{Step $1$}:
	At the start, the GPR-PES is built with the energy and gradient information
	from the starting point. We find the minimum on the GPR-PES 
	shown by the star symbol.
	This is our next estimate for the PES minimum, and we
	calculate the energy and gradient of the PES
	at that position,
	indicated by the arrow.\\
	\item \emph{Step $2$}: 
	After evaluating the energy and gradient at the estimate
	from the last step, we build up the next GPR-PES with now
	two training points. 
	The new minimum of the GPR-PES, however, 
	leads us in the same direction as in the last step.
	Therefore, we scale up the estimated step size in the 
	overshooting procedures described above.
	The overshooting to a more distant point is indicated by
	the tilted arrow, which points to the next estimate 
	at which we 
	calculate the energy and the gradient.
	If the estimated step size is now larger
	than the externally set maximum step size $s_{\mathrm{max}}$,
	we scale the step down to a step size
	of $s_{\mathrm{max}}$.\\
	\item \emph{Step $3$}:
	The minimum on the new GPR-PES with now 
	three training points, leads to a step
	in the opposite direction of the last step.
	Therefore, no overshooting is performed.\\
	\item \emph{Step $4$}:
	The next estimate for the minimum is close enough 
	to the last
	estimated minimum, 
	so the convergence criteria for the step size 
	are satisfied.
	Calculating the gradient at 
	estimate $4$, will also show that
	the convergence criteria 
	for the gradient are satisfied.
	The optimizer is completely converged.
	\end{itemize}		
	The limitation of the step size, called
	$s_{\mathrm{max}}$, lies
	roughly between $0.5$ and $1\; a.u.$
	This prevents the overshooting process from
	shooting in a region outside of the 
	domain in which the chosen
	electronic structure calculations
	are valid.
						
	\subsubsection{Parameters}
	\label{sec::parameters}
	For all results presented in this paper we used the 
	Mat\'{e}rn covariance function
	of \eqref{eq::M_Kernel}.
	We chose the parameter $\sigma_\text{f}=1$ since it does not influence
	the result. 
	The only other parameter in the covariance function
	is $l$.
	We chose $l=20$ at the beginning of the optimization.
	Then we chose a dynamic approach:
	Every step along the optimization on which the gradient has
	become larger, instead of smaller, we increase $1/l^2$ 
	by $10\%$ of its current value.
	This leads to a shrinking characteristic length scale along
	the optimization,
	which means that the steps predicted by the GPR optimizer
	will become smaller, the training points closer.
	A smaller characteristic length scale is also advisable
	towards the end of an optimization procedure, since we often need
	more careful steps as we approach the minimum.
	The noise parameters $\sigma_\text{e}$ and $\sigma_\text{g}$, see equations 
	(\ref{eq::cov_mat_elements}) and (\ref{eq:KM_firstDeriv}),
	are chosen to be $\sigma_\text{e}=\sigma_\text{g}=10^{-7}$
	which is a compromise between the smallest possible
	value, and numerical stability we found by
	using cross validation on several systems.
	It may change for different electronic structure methods
	or codes.
	We also found that the Mat\'{e}rn covariance function
	is rather insensitive to changes of the $\sigma$ parameters.
	One can also use the 
	\emph{maximum likelihood principle}\cite{rasmussen2006gaussian}
	to optimize the parameters. 
	Still, in our test cases we found it
	to be less useful than cross validation,
	and less successful than our dynamic approach.\\
	
	We also include an offset in the form of the prior mean function,
	see \eqref{eq::GPprediction}.
	Away from any training point the GPR-PES will slowly converge
	to that value.
	This is a fact we exploit in our optimizer:
	The prior mean is chosen to be a constant that is
	much higher than the energy
	values observed in the system.
	This will restrict the optimization to a reasonable
	area around the observed training points, and 
	guarantees that a minimum on the GPR-PES can be found
	at any time.
	The prior mean for our minimization procedure is chosen
	to be 
	\begin{equation}
	\label{eq::priorMeanEnergyOffset}
	E_{\text{mean}}= \max\limits_{i}E_i + 10
	\end{equation}
    The value of $E_{\text{mean}}$ 
	can change with
	an increasing amount of training points and is reevaluated,
	if new training points are added to the GPR-PES.
		
	Just like in L-BFGS optimizations, the maximum step size
    $s_{\mathrm{max}}$ is the only parameter which has to be specified by
    the user.
    \ad{The parameters in the overshooting schemes
    were chosen to provide reasonable 
    performance on the Baker test set we show in
    Section~\ref{sec::applications}.
    The sensitivity of the performance on these parameters 
    is small.}
	
	\subsubsection{Multi-level GPR}
	\label{sec::multiLevelGPR}
	In GPR 
	one needs to solve a linear system with 
	the covariance matrix.
	If we include gradient information, the size of the covariance matrix 
	is approximately $Nd\times Nd$, see \eqref{eq:KM_firstDeriv}. 
	$N$ is the number of points at which we have gradient and energy information,
	$d$ is the number of degrees of freedom of the system. 
	The size of the matrix results from the fact that
	every entry of the gradient is
	considered to be a new training point.
	The solution of the linear system with 
	the covariance matrix 
	is carried out via Cholesky decomposition.
	Therefore, the required CPU time to solve the linear system
	scales with $\mathcal{O}\left(N^3d^3\right)$. 
	In high dimensional systems and long optimization runs
	this can become computationally more demanding than density functional
        theory (DFT) calculations.
	To overcome this problem one can restrict the GPR to, e.g., the last
	$50$ training points which leads to a formal scaling of $\mathcal{O}\left(d^3\right)$.
	That is independent of the length of the optimization history.\\
	
	We found that it is more efficient to use 
	the neglected training points
	to build up another GP. 
	The other training points are then used to
	minimize the error of that GP.
	In our code we build up a hierarchical multi-level GP:
	As soon as the number of training points reaches $N_{\mathrm{max}}$,
	we take the oldest $m$ of them to build a GP, called $GP_1$. 
	The remaining $N_{\mathrm{max}}-m$ 
	elements of the training set are used to learn the error of $GP_1$
	to give a new surface $GP_0$:
	We use $GP_1$ as the prior mean function for $GP_0$.
	Along the optimization procedure new training points are added to $GP_0$.
	As soon as their number reaches $N_{\mathrm{max}}$ again we rename $GP_1$ to 
	$GP_2$. We use the $m$ oldest training points of $GP_0$ to build up a new $GP_1$, that
	uses $GP_2$ as its prior mean function. 
	The remaining $N_{\mathrm{max}}-m$ training points of $GP_0$ are used to
	build a new surface $GP_0$, that uses $GP_1$ as its prior 
	mean function.
	This process is repeated by increasing the number of levels
	as soon as the number of training points in $GP_0$ reaches
	$N_{\mathrm{max}}$.
	The last $GP_q$, with the highest $q$, uses the usual offset of
	\eqref{eq::priorMeanEnergyOffset}.
	Only the $E_i$ included in this $GP_q$ are used
	to calculate this offset.
	
	The multi-level approach decreases the accuracy of the 
	regression compared to full GPR near the training points
	that are not included in $GP_0$.
	However, the most relevant information for an optimizer is most likely
	encoded in the last few training points 
	which are still included in $GP_0$. 
	We found it to be sufficient to set $N_{\mathrm{max}}=60$, and $m=10$,
	to keep relatively good performance whilst requiring much lower computational
	cost than the DFT method. 
	We set these values in all the presented test cases
	in this paper.
	
	\section{Applications}
	\label{sec::applications}
\begin{figure}
   		\begin{center}
   			\includegraphics[width=8cm]{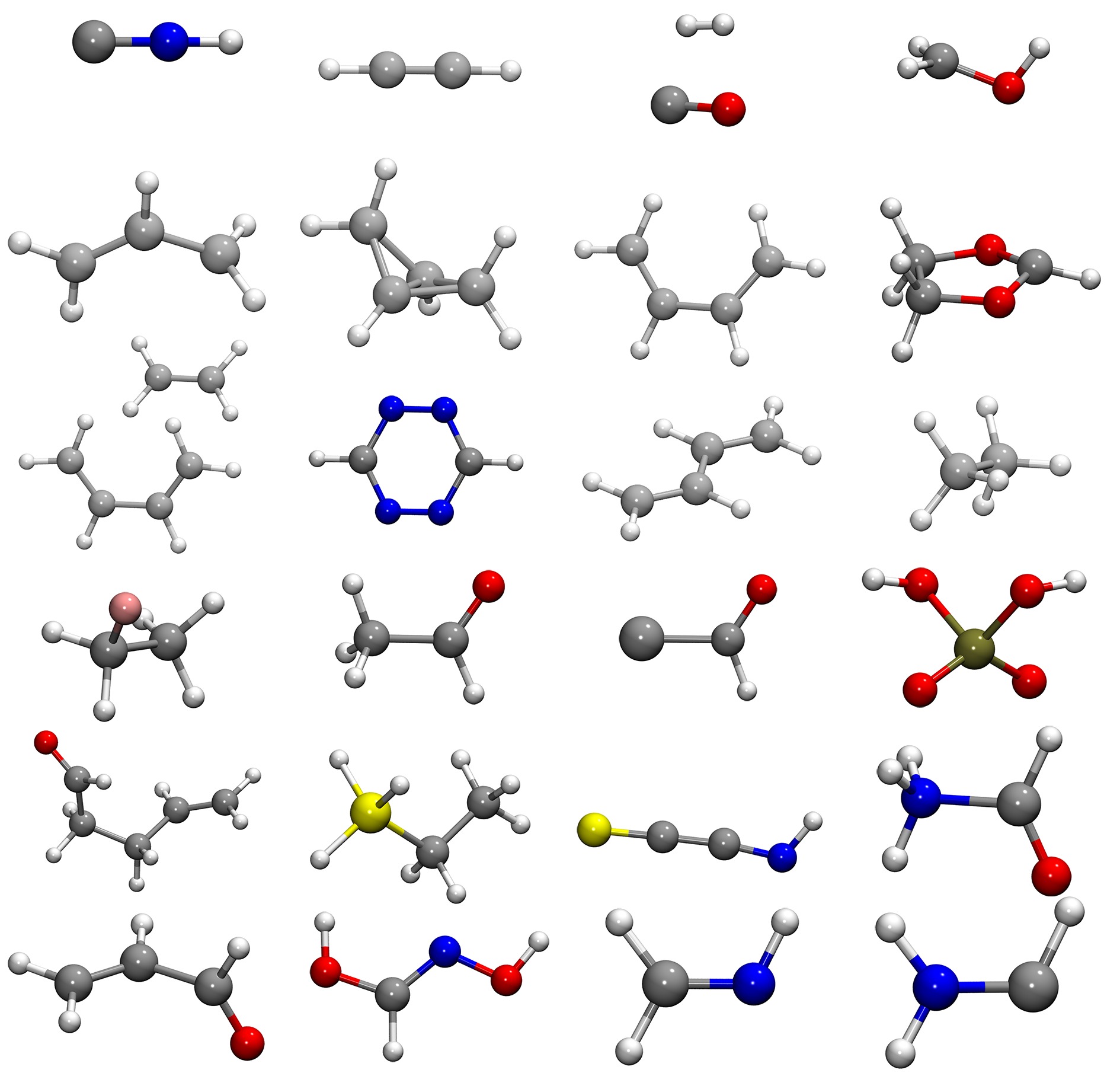}
   			\caption{\ad{The minima of the test systems suggested by Baker.\cite{Baker}
   					The systems are ordered lines first, columns second.
   					The structures with ID $23$ and $24$ 
   					are the same and only shown once.}}
			\label{fig::Baker_pictures}
  		\end{center}
  	\end{figure}

	We apply our optimization algorithm to several test cases.
	We chose a set of $25$ test systems suggested by Baker.\cite{Baker}
	The starting points of the optimization 
	were chosen following Ref. \citenum{Baker} close to a transition state
        \ad{on the Hartree--Fock level.} 
	In contrast to Ref. \citenum{Baker} we use the semi-empirical
	AM1\cite{dewar1985development} method for the electronic
	structure calculations. \ad{The resulting minimum structures are shown
          in \figref{fig::Baker_pictures}.}	
These tests are in the following referred to
        with IDs $1$ to $25$. \ad{Note that the structures with ID 23 and 24
          start at different geometries, but end up in the same minimum.}
    
\begin{figure}
		\begin{center}
		\includegraphics[width=8cm]{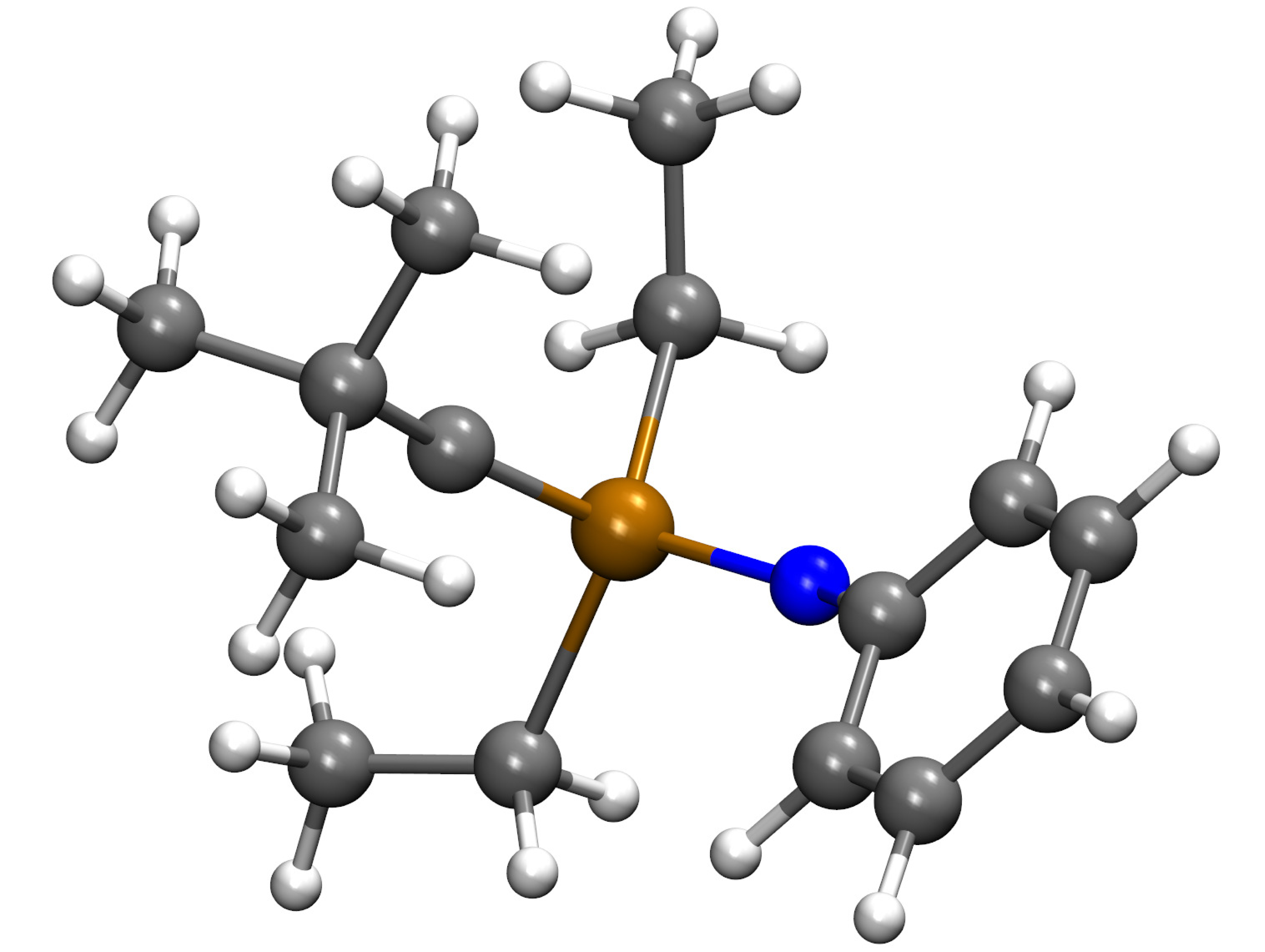}
		\caption{The minimum structure of the molybdenum 
				amidato bisalkyl alkylidyne complex
				found by our GPR optimizer in the
				run with ID $26$. Molybdenum is
				depicted in golden brown, nitrogen
				in blue, carbon in grey and 
				hydrogen in white.
			}
			\label{fig::molybdenumComplex}
		\end{center}
	\end{figure}

	Additionally, we set up a more realistic test case:
	We use a part of a previously investigated molybdenum 
	amidato bisalkyl alkylidyne complex\cite{Molybdenum}
	that includes $41$ atoms, see \figref{fig::molybdenumComplex}. 
	Electronic structure calculations are carried out 
	with the BP86 functional\cite{bec88,perdew1986} in the def2-SVP basis set.\cite{wei05}
	We give the optimization runs on this
	\emph{molybdenum system} the
	IDs $26$, $27$, and $28$.
	We chose a different starting point for the optimization
	in run $27$ than in the other two.
	Runs $26$ and $28$ begin at
	the same starting point and
	only differ in the chosen
	convergence criteria:
	The convergence criteria were chosen according to equations
	($\ref{eq::convCriteria1}-\ref{eq::convCriteria4}$)
	with $\delta = 4.5\cdot 10^{-4}$ 
	for the run on the 
	molybdenum system with IDs $26$ and $27$.
	We chose the stricter criterion $\delta = 1\cdot 10^{-4}$ 
	for the run on the molybdenum system with ID $28$,
	and we set $\delta = 3\cdot 10^{-4}$ 
	for the \emph{Baker systems} with IDs $1$ to $25$.\\
			
	The maximum step size was set to $5\; a.u.$ (never reached) for
	L-BFGS since it yields the best performance like that.
	The maximum step size for our GPR optimizer
	was set to $0.5\; a.u.$ for the Baker systems, and 
	$1\; a.u.$ for the molybdenum system runs. 
	The number of steps in the L-BFGS memory
	is chosen to be $50$ for the molybdenum system,
	and equal to the number of dimensions
	in the Baker systems.
	\ad{The L-BFGS optimizer in \dl employs
	a variable trust radius based on energy decrease.\cite{dlfind}}\\
	
	We performed all the presented calculations in 
	Cartesian coordinates.
	The GPR optimizer is in principle able to handle
	other coordinates, but the adaptations
	of the algorithm needed to perform well with these
	is not trivial. Especially the optimal
	length scale parameter $l$
	may be different in every dimension.
	Our implementation is not able to do that yet.

	\subsection{Comparison to L-BFGS}
	\label{sec::comparison_to_lbfgs}
	
	Rigorously proving the convergence order of our optimizer
	is rather difficult, if not impossible. 
	Instead we present a comparison 
	to the super-linearly converging
	L-BFGS algorithm that is implemented in \dl.
	In \tabref{tab::nSteps} we show 
	a comparison between the GPR and the L-BFGS optimizer in
	\dl for the different test systems.
	\ad{We compare the number of steps both optimizers
		take until convergence and we compare 
		the obtained minima
		according to their energy and the RMSD value of their geometries.
		In the supplementary material we show a 
		further comparison to the steepest descent and
		the conjugate gradient methods, which perform much 
		worse than L-BFGS.
		We also show the optimization runs
		on the Baker test set using DFT instead of AM1.
		}\\
	
	\begin{table}[htbp!]
		\caption{
			A comparison of the L-BFGS and the GPR
			optimizer in our test systems, 
			sorted by the number of dimensions $d$ in the system.
			\ad{The number of steps required until convergence is given for
			L-BFGS and the GPR optimizer, $\Delta \text{steps}$ is their
                        difference, 
			$\Delta \text{energy}$ is the energy difference 
			between the minima in Hartree, RMSD denotes the 
			root-mean-square deviation of their atomic positions 
			in {\AA}ng.}
			}
	{\setlength{\extrarowheight}{2pt}
		\bgroup
		\begin{tabular}{rrrrrlr}
			\hline
			\hline
                 & \multicolumn{2}{c}{steps}\\[-1.5ex]
			\multicolumn{1}{c}{$d$} & 
			\multicolumn{1}{c}{GPR} & 
			\multicolumn{1}{c}{L-BFGS} & 
			\multicolumn{1}{c}{$\Delta \text{steps}$} & 
			\multicolumn{1}{c}{$\Delta \text{energy}$} & 
			\multicolumn{1}{c}{RMSD} & 
			\multicolumn{1}{c}{ID} \\[-0.5ex] 
\hline
$123$ & $105$ & $233$ & $128$ & $-5.67\cdot 10^{-04}$ & $2.71\cdot 10^{-01}$ & $26$ \\[-1.5ex] 
$123$ & $106$ & $206$ & $100$ & $-6.16\cdot 10^{-04}$ & $2.86\cdot 10^{-01}$ & $27$ \\[-1.5ex] 
$123$ & $173$ & $418$ & $245$ & $-3.36\cdot 10^{-04}$ & $2.51\cdot 10^{-01}$ & $28$ \\[-1.5ex] 
$48$ & $77$ & $83$ & $6$ & $-1.15\cdot 10^{-04}$ & $1.37\cdot 10^{0}$ & $9$ \\[-1.5ex] 
$42$ & $109$ & $104$ & $-5$ & $-2.20\cdot 10^{-05}$ & $6.38\cdot 10^{-02}$ & $17$ \\[-1.5ex] 
$33$ & $29$ & $39$ & $10$ & $-3.68\cdot 10^{-07}$ & $1.44\cdot 10^{-03}$ & $18$ \\[-1.5ex] 
$30$ & $26$ & $27$ & $1$ & $1.93\cdot 10^{-06}$ & $2.34\cdot 10^{-03}$ & $6$ \\[-1.5ex] 
$30$ & $68$ & $62$ & $-6$ & $2.16\cdot 10^{-08}$ & $2.58\cdot 10^{-03}$ & $7$ \\[-1.5ex] 
$30$ & $38$ & $42$ & $4$ & $-4.17\cdot 10^{-08}$ & $3.08\cdot 10^{-04}$ & $8$ \\[-1.5ex] 
$30$ & $34$ & $37$ & $3$ & $-1.68\cdot 10^{-08}$ & $1.12\cdot 10^{-03}$ & $11$ \\[-1.5ex] 
$24$ & $29$ & $31$ & $2$ & $2.29\cdot 10^{-07}$ & $9.15\cdot 10^{-04}$ & $5$ \\[-1.5ex] 
$24$ & $13$ & $15$ & $2$ & $-6.40\cdot 10^{-09}$ & $6.60\cdot 10^{-05}$ & $10$ \\[-1.5ex] 
$24$ & $15$ & $23$ & $8$ & $1.67\cdot 10^{-08}$ & $4.45\cdot 10^{-02}$ & $12$ \\[-1.5ex] 
$24$ & $14$ & $19$ & $5$ & $1.09\cdot 10^{-07}$ & $3.65\cdot 10^{-03}$ & $13$ \\[-1.5ex] 
$24$ & $19$ & $22$ & $3$ & $-9.58\cdot 10^{-08}$ & $3.83\cdot 10^{-04}$ & $21$ \\[-1.5ex] 
$21$ & $16$ & $23$ & $7$ & $2.52\cdot 10^{-07}$ & $3.36\cdot 10^{-04}$ & $14$ \\[-1.5ex] 
$21$ & $30$ & $80$ & $50$ & $-7.08\cdot 10^{-02}$ & $8.76\cdot 10^{-01}$ & $16$ \\[-1.5ex] 
$21$ & $22$ & $25$ & $3$ & $-1.45\cdot 10^{-08}$ & $2.77\cdot 10^{-02}$ & $20$ \\[-1.5ex] 
$21$ & $19$ & $22$ & $3$ & $1.75\cdot 10^{-07}$ & $7.88\cdot 10^{-03}$ & $22$ \\[-1.5ex] 
$15$ & $12$ & $14$ & $2$ & $8.47\cdot 10^{-09}$ & $8.27\cdot 10^{-04}$ & $4$ \\[-1.5ex] 
$15$ & $13$ & $26$ & $13$ & $-3.50\cdot 10^{-06}$ & $1.87\cdot 10^{-02}$ & $19$ \\[-1.5ex] 
$15$ & $14$ & $19$ & $5$ & $-1.04\cdot 10^{-07}$ & $2.82\cdot 10^{-04}$ & $23$ \\[-1.5ex] 
$15$ & $19$ & $23$ & $4$ & $-5.16\cdot 10^{-08}$ & $2.16\cdot 10^{-04}$ & $24$ \\[-1.5ex] 
$15$ & $25$ & $29$ & $4$ & $1.19\cdot 10^{-07}$ & $3.52\cdot 10^{-04}$ & $25$ \\[-1.5ex] 
$12$ & $16$ & $26$ & $10$ & $5.02\cdot 10^{-08}$ & $2.05\cdot 10^{-04}$ & $2$ \\[-1.5ex] 
$12$ & $18$ & $46$ & $28$ & $-2.18\cdot 10^{-04}$ & $3.62\cdot 10^{-01}$ & $3$ \\[-1.5ex] 
$12$ & $12$ & $13$ & $1$ & $6.20\cdot 10^{-10}$ & $3.50\cdot 10^{-02}$ & $15$ \\[-1.5ex] 
$9$ & $15$ & $19$ & $4$ & $1.00\cdot 10^{-11}$ & $3.66\cdot 10^{-02}$ & $1$ \\[-1ex] 
\hline
\hline
		\end{tabular}
		\egroup
	}
		\label{tab::nSteps}
	\end{table}

	The GPR optimizer yields quite good results. 
	In most cases it is 
	faster than the L-BFGS optimizer.
	\ad{Some qualitative differences can be observed. In the case of
          system $16$ the GPR optimizer finds a different minimum than L-BFGS.
          One of these two different minima represents the reactant, the other
          one the product of this system. This happens because the
          optimization starts in the vicinity of the transition
          structure. System $3$ and $9$ show high RMSD values between the
          structures while their energy differences vanish: These are
          bimolecular reactions in which the minimum region of the separated
          molecules is flat using AM1.}  In the case of the molybdenum system
        with ID $26$ and $27$ the GPR optimizer finds a minimum that is higher
        in energy, and a little closer to the starting point.  The minima look
        similar with slightly different torsions in the aliphatic groups.  For
        the molybdenum system with ID $28$ we applied stricter convergence
        criteria: the GPR optimizer is significantly faster, but again finds a
        minimum that is a little higher in energy.  Towards the end of the
        optimization the convergence of L-BFGS is mostly hindered by larger
        predicted step sizes.  \ad{All obtained minima look chemically
          plausible and (except for system $16$) similar comparing the results
          of L-BFGS and the GPR optimizer.}\\

\begin{figure}
			\begin{center}
				\includegraphics[width=8cm]{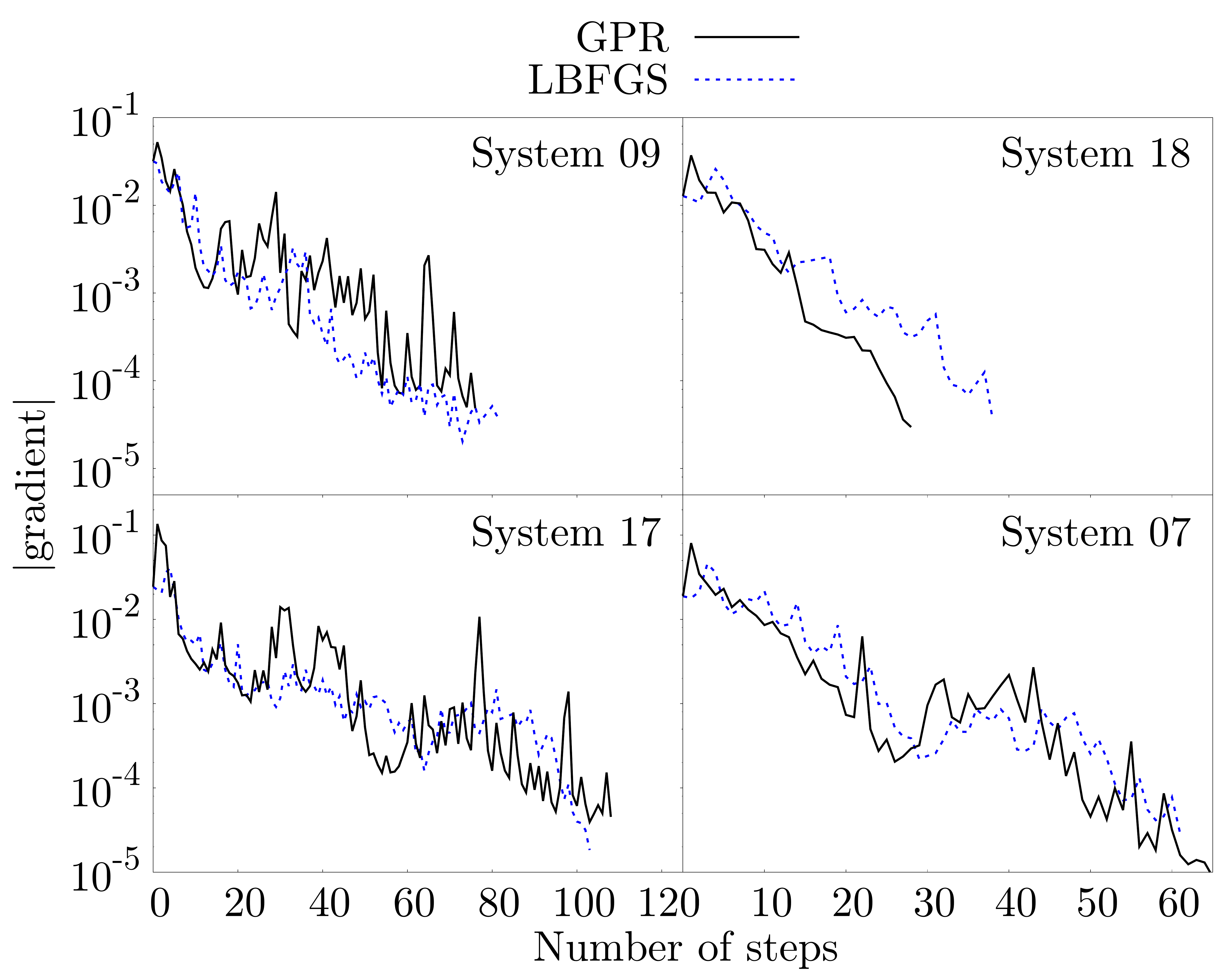}
				\caption{The Euclidean norm of the gradient
					with respect
					to the number of steps taken by the 
					L-BFGS and the GPR optimizer
					in $4$ of the biggest 
					systems from the Baker test set.	
				}
				\label{fig::convergence_baker}
			\end{center}
		\end{figure}

\begin{figure}
			\begin{center}
				\includegraphics[width=8cm]{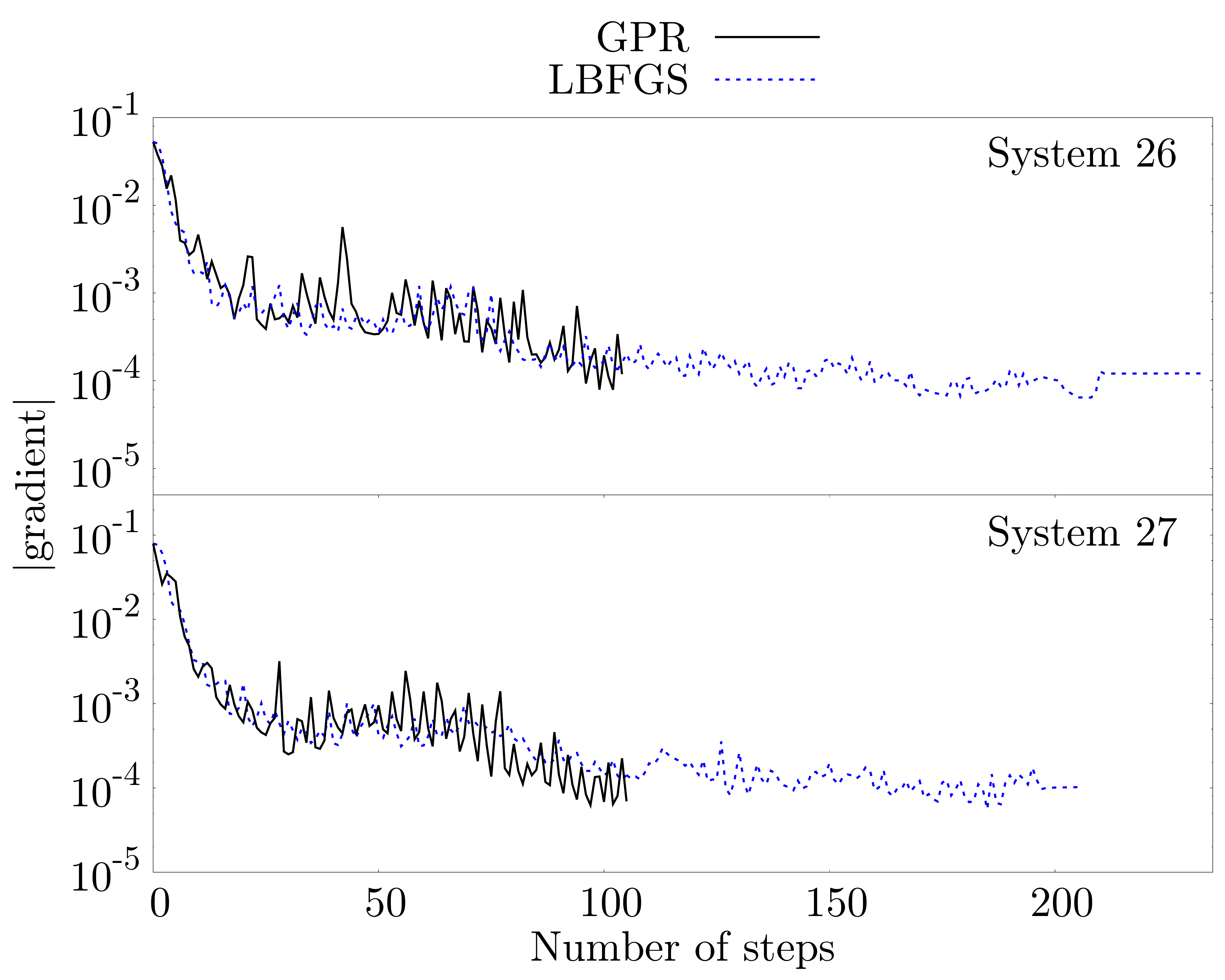}
				\caption{The Euclidean norm of the gradient
					with respect
					to the number of steps taken by the 
					L-BFGS and the GPR optimizer
					in the molybdenum system with two different
					starting points.
				}
				\label{fig::convergence_molybdenum}
			\end{center}
		\end{figure}

	For some of the test cases,
	we show the convergence rates as the
	Euclidean norm of the gradient versus 
	the number of steps taken:
	We depict the convergence rate of four of the biggest test cases
	from the Baker test set in \figref{fig::convergence_baker}.
	We also show the convergence rates of the 
	runs on the molybdenum system with IDs $26$ and $27$,
	that have two different starting points
	in \figref{fig::convergence_molybdenum}.
	The higher fluctuations of the GPR optimizer compared to the L-BFGS
	optimizer are due to the overshooting procedures, see Sections~\ref{sec::gpr_optimizer} and~\ref{sec::sepDimOvershooting}.
	They are not necessarily a sign for bad performance of the
	algorithm. The high overshooting is intentional,
	and decreases the overall number of steps needed in almost
	all cases.
	The convergence criteria concerning the step size are usually
	fulfilled later than the ones concerning the
	gradient. This explains the large amount of steps  
	that L-BFGS uses in the molybdenum system,
	although, the convergence criterion for the
	gradients are already met.
	Also the L-BFGS optimizer discards a lot of 
	steps when the energy increases along the optimization,
	especially at the end. 
	This can be seen when the norm of the 
	gradient stays constant for some time:
	No actual step is taken.
	The discarding of steps is not necessary for the GPR optimizer.
		
	\subsection{Timing}
	\label{sec::timing}
	The GPR optimizer is more demanding 
	in terms of computational power per step.
	The Cholesky decomposition 
	of the matrix scales with the third power
	of the dimension, and the amount of training points.
	We formally eliminate the scaling with respect
	to the number of training points by the 
	GPR multi-level approach, see Section~\ref{sec::multiLevelGPR}.
	Nevertheless, the optimizer takes more time per step
	than the L-BFGS optimizer.
	This can best be seen in our biggest test case, 
	the molybdenum system (IDs $26$ to $28$):
	The GPR optimizer
	took about $10\%$ of the overall computational time.
	The L-BFGS optimizer procedures take around $1\%$.
	Omitting the multi-level approach 
	will result in only slightly faster
	convergence of the GPR optimizer. 
	This is in terms of number of 
	optimization steps. 
	Nevertheless, the time needed for the optimizer procedures
	will be almost equal to the time consumption of the DFT calculations.
	The additional overhead of approximately
	$9\%$ using the multi-level approach 
	is easily compensated 
	by the faster convergence in our test case. 
	In the run with ID $26$,
	the GPR optimization took around $48$ minutes in total,
	while the L-BFGS optimization took around $88$ minutes. 
	The runs were all performed
	on an Intel i5-4570 quad-core CPU.
	If we chose more accurate electronic structure methods 
	than the DFT method we used here,
	and look at correspondingly smaller systems,
	we expect only a negligible overhead from the GPR optimizer.
		
	\section{Discussion\label{sec::discussion}}
	
        We presented an algorithm to use GPR to find minima on the PES
        using Cartesian coordinates.  Different coordinate systems
        that intrinsically incorporate translational and rotational
        invariance, and invariance to permutations of identical atoms
        can improve the efficiency of machine learning methods.
        \cite{ramakrishnan2017,hansen2015interaction,
          Behler2016,ramakrishnan2017,Bartok2013} The incorporation of
        those would possibly require different characteristic length
        scale in the covariance function for the different
        coordinates. We prefer Cartesian coordinates because of their
        simplicity.
	
	We tested only two covariance functions for our algorithm.
	Combinations of those, or inclusion of
	other covariance functions might improve the efficiency of the algorithm.
	

	Generally, it is known that machine learning methods like GPR
        are best in learning small errors of the energy between
        different methods, rather than learning the exact energy
        directly.  Therefore, one could use a fast estimate of the PES
        using e.g. a simple force field method as a prior of the GPR
        just like we did with the higher levels in our multi-level
        approach.  This could further improve the performance, and the
        additional overhead may pay off for high-precision
        calculations.
	
	\ad{In higher dimensional systems GPR is less
          promising since the covariance matrix can become
          sparse.\cite{Habershon_Tew}  In our
          investigated systems no such indications were found.  Even
          in the $123$-dimensional molybdenum system GPR performs
          well.  On the other hand, due to the high scaling of its
          computational effort with the number of dimensions GPR is 
          recommended mainly for small systems in any case.  Furthermore, the
          fixed number of $60$ training points in the last level of
          our multi-level
          approach may have to be increased for larger systems.}\\
	
	The overall result of our benchmark indicates
	a good performance of the new GPR optimizer.
	The big advantage of the GPR optimizer compared
	to traditional optimizers is that it can do comparably large steps:
	If a step overshoots the minimum, or even yields a
	higher energy structure than before, 
	this generally improves the performance of the optimizer.
	It can easily use the obtained information from an
	overestimated step to build a more conservative step later.
	This results in the possibility of doing
	large steps without compromising the efficiency
	of the optimizer. 
	The possibility to do large steps like this 
	can speed up the optimization procedure
	as we have seen in our largest test case.	
	
	\section{Conclusions}
	\label{sec::conclusion}
	We presented a new optimizer for finding minima on potential
        energy surfaces using GPR.  It outperforms the
        super-linearly converging L-BFGS optimizer.  Using only one
        external parameter, a step size limit, it is easy to apply.
        It is included in \dl and hence, usable in \chemshell.  We
        will further investigate possibilities for improvement, some
        mentioned in Section~\ref{sec::discussion},
	and try to extend the optimizer to saddle point searches.\\
	
	
        \textbf{Supporting Information:} Derivations for including second
        derivatives, results of the Baker set for conjugate gradient
        and steepest descent optimizers, as well as with DFT, are provided as supporting information.
	
	\begin{acknowledgments} We thank Marc Toussaint for
          stimulating discussions. This work was financially supported by the
          European Union's Horizon 2020 research and innovation programme
          (grant agreement No. 646717, TUNNELCHEM) and the German Research
          Foundation (DFG) through the Cluster of Excellence in Simulation
          Technology (EXC 310/2) at the University of Stuttgart.
	\end{acknowledgments}
	
	\bibliography{literature}

\begin{thebibliography}{35}%
\makeatletter
\providecommand \@ifxundefined [1]{%
 \@ifx{#1\undefined}
}%
\providecommand \@ifnum [1]{%
 \ifnum #1\expandafter \@firstoftwo
 \else \expandafter \@secondoftwo
 \fi
}%
\providecommand \@ifx [1]{%
 \ifx #1\expandafter \@firstoftwo
 \else \expandafter \@secondoftwo
 \fi
}%
\providecommand \natexlab [1]{#1}%
\providecommand \enquote  [1]{``#1''}%
\providecommand \bibnamefont  [1]{#1}%
\providecommand \bibfnamefont [1]{#1}%
\providecommand \citenamefont [1]{#1}%
\providecommand \href@noop [0]{\@secondoftwo}%
\providecommand \href [0]{\begingroup \@sanitize@url \@href}%
\providecommand \@href[1]{\@@startlink{#1}\@@href}%
\providecommand \@@href[1]{\endgroup#1\@@endlink}%
\providecommand \@sanitize@url [0]{\catcode `\\12\catcode `\$12\catcode
  `\&12\catcode `\#12\catcode `\^12\catcode `\_12\catcode `\%12\relax}%
\providecommand \@@startlink[1]{}%
\providecommand \@@endlink[0]{}%
\providecommand \url  [0]{\begingroup\@sanitize@url \@url }%
\providecommand \@url [1]{\endgroup\@href {#1}{\urlprefix }}%
\providecommand \urlprefix  [0]{URL }%
\providecommand \Eprint [0]{\href }%
\providecommand \doibase [0]{http://dx.doi.org/}%
\providecommand \selectlanguage [0]{\@gobble}%
\providecommand \bibinfo  [0]{\@secondoftwo}%
\providecommand \bibfield  [0]{\@secondoftwo}%
\providecommand \translation [1]{[#1]}%
\providecommand \BibitemOpen [0]{}%
\providecommand \bibitemStop [0]{}%
\providecommand \bibitemNoStop [0]{.\EOS\space}%
\providecommand \EOS [0]{\spacefactor3000\relax}%
\providecommand \BibitemShut  [1]{\csname bibitem#1\endcsname}%
\let\auto@bib@innerbib\@empty
\bibitem [{\citenamefont {Hestenes}\ and\ \citenamefont
  {Stiefel}(1952)}]{CG_Method}%
  \BibitemOpen
  \bibfield  {author} {\bibinfo {author} {\bibfnamefont {M.~R.}\ \bibnamefont
  {Hestenes}}\ and\ \bibinfo {author} {\bibfnamefont {E.}~\bibnamefont
  {Stiefel}},\ }\href@noop {} {\bibfield  {journal} {\bibinfo  {journal} {J.
  Res. Natl. Bur. Stand.}\ }\textbf {\bibinfo {volume} {49}},\ \bibinfo {pages}
  {409} (\bibinfo {year} {1952})}\BibitemShut {NoStop}%
\bibitem [{\citenamefont {Broyden}(1970)}]{Broyden:1970:CCDa}%
  \BibitemOpen
  \bibfield  {author} {\bibinfo {author} {\bibfnamefont {C.~G.}\ \bibnamefont
  {Broyden}},\ }\href@noop {} {\bibfield  {journal} {\bibinfo  {journal} {J.
  Inst. Maths. Appl.}\ }\textbf {\bibinfo {volume} {6}},\ \bibinfo {pages} {76}
  (\bibinfo {year} {1970})}\BibitemShut {NoStop}%
\bibitem [{\citenamefont {Fletcher}(1970)}]{fletcher1970new}%
  \BibitemOpen
  \bibfield  {author} {\bibinfo {author} {\bibfnamefont {R.}~\bibnamefont
  {Fletcher}},\ }\href@noop {} {\bibfield  {journal} {\bibinfo  {journal}
  {Comp. J.}\ }\textbf {\bibinfo {volume} {13}},\ \bibinfo {pages} {317}
  (\bibinfo {year} {1970})}\BibitemShut {NoStop}%
\bibitem [{\citenamefont {Goldfarb}(1970)}]{goldfarb1970family}%
  \BibitemOpen
  \bibfield  {author} {\bibinfo {author} {\bibfnamefont {D.}~\bibnamefont
  {Goldfarb}},\ }\href@noop {} {\bibfield  {journal} {\bibinfo  {journal}
  {Math. Comp.}\ }\textbf {\bibinfo {volume} {24}},\ \bibinfo {pages} {23}
  (\bibinfo {year} {1970})}\BibitemShut {NoStop}%
\bibitem [{\citenamefont {Shanno}(1970)}]{shanno1970conditioning}%
  \BibitemOpen
  \bibfield  {author} {\bibinfo {author} {\bibfnamefont {D.~F.}\ \bibnamefont
  {Shanno}},\ }\href@noop {} {\bibfield  {journal} {\bibinfo  {journal} {Math.
  Comp.}\ }\textbf {\bibinfo {volume} {24}},\ \bibinfo {pages} {647} (\bibinfo
  {year} {1970})}\BibitemShut {NoStop}%
\bibitem [{\citenamefont {Fletcher}(1980)}]{fletcher1980practical}%
  \BibitemOpen
  \bibfield  {author} {\bibinfo {author} {\bibfnamefont {R.}~\bibnamefont
  {Fletcher}},\ }\href@noop {} {\emph {\bibinfo {title} {Practical methods of
  optimization}}}\ (\bibinfo  {publisher} {Wiley, New York},\ \bibinfo {year}
  {1980})\BibitemShut {NoStop}%
\bibitem [{\citenamefont {Liu}\ and\ \citenamefont {Nocedal}(1989)}]{Liu1989}%
  \BibitemOpen
  \bibfield  {author} {\bibinfo {author} {\bibfnamefont {D.~C.}\ \bibnamefont
  {Liu}}\ and\ \bibinfo {author} {\bibfnamefont {J.}~\bibnamefont {Nocedal}},\
  }\href@noop {} {\bibfield  {journal} {\bibinfo  {journal} {Math. Program.}\
  }\textbf {\bibinfo {volume} {45}},\ \bibinfo {pages} {503} (\bibinfo {year}
  {1989})}\BibitemShut {NoStop}%
\bibitem [{\citenamefont {Nocedal}(1980)}]{Nocedal1980}%
  \BibitemOpen
  \bibfield  {author} {\bibinfo {author} {\bibfnamefont {J.}~\bibnamefont
  {Nocedal}},\ }\href@noop {} {\bibfield  {journal} {\bibinfo  {journal} {Math.
  Comput.}\ }\textbf {\bibinfo {volume} {35}},\ \bibinfo {pages} {773}
  (\bibinfo {year} {1980})}\BibitemShut {NoStop}%
\bibitem [{\citenamefont {Behler}(2016)}]{Behler2016}%
  \BibitemOpen
  \bibfield  {author} {\bibinfo {author} {\bibfnamefont {J.}~\bibnamefont
  {Behler}},\ }\href {\doibase 10.1063/1.4966192} {\bibfield  {journal}
  {\bibinfo  {journal} {J. Chem. Phys.}\ }\textbf {\bibinfo {volume} {145}},\
  \bibinfo {pages} {170901} (\bibinfo {year} {2016})}\BibitemShut {NoStop}%
\bibitem [{\citenamefont {Alborzpour}, \citenamefont {Tew},\ and\ \citenamefont
  {Habershon}(2016)}]{Habershon_Tew}%
  \BibitemOpen
  \bibfield  {author} {\bibinfo {author} {\bibfnamefont {J.~P.}\ \bibnamefont
  {Alborzpour}}, \bibinfo {author} {\bibfnamefont {D.~P.}\ \bibnamefont {Tew}},
  \ and\ \bibinfo {author} {\bibfnamefont {S.}~\bibnamefont {Habershon}},\
  }\href {\doibase 10.1063/1.4964902} {\bibfield  {journal} {\bibinfo
  {journal} {J. Chem. Phys.}\ }\textbf {\bibinfo {volume} {145}},\ \bibinfo
  {pages} {174112} (\bibinfo {year} {2016})}\BibitemShut {NoStop}%
\bibitem [{\citenamefont {Bart\'ok}, \citenamefont {Kondor},\ and\
  \citenamefont {Cs\'anyi}(2013)}]{Bartok2013}%
  \BibitemOpen
  \bibfield  {author} {\bibinfo {author} {\bibfnamefont {A.~P.}\ \bibnamefont
  {Bart\'ok}}, \bibinfo {author} {\bibfnamefont {R.}~\bibnamefont {Kondor}}, \
  and\ \bibinfo {author} {\bibfnamefont {G.}~\bibnamefont {Cs\'anyi}},\ }\href
  {\doibase 10.1103/PhysRevB.87.184115} {\bibfield  {journal} {\bibinfo
  {journal} {Phys. Rev. B}\ }\textbf {\bibinfo {volume} {87}},\ \bibinfo
  {pages} {184115} (\bibinfo {year} {2013})}\BibitemShut {NoStop}%
\bibitem [{\citenamefont {Ramakrishnan}\ and\ \citenamefont {von
  Lilienfeld}(2017)}]{ramakrishnan2017}%
  \BibitemOpen
  \bibfield  {author} {\bibinfo {author} {\bibfnamefont {R.}~\bibnamefont
  {Ramakrishnan}}\ and\ \bibinfo {author} {\bibfnamefont {O.~A.}\ \bibnamefont
  {von Lilienfeld}},\ }\enquote {\bibinfo {title} {Machine learning, quantum
  chemistry, and chemical space},}\ in\ \href {\doibase
  10.1002/9781119356059.ch5} {\emph {\bibinfo {booktitle} {Reviews in
  Computational Chemistry}}}\ (\bibinfo  {publisher} {JWS},\ \bibinfo {year}
  {2017})\ pp.\ \bibinfo {pages} {225--256}\BibitemShut {NoStop}%
\bibitem [{\citenamefont {Mills}\ and\ \citenamefont
  {J\'{o}nsson}(1994)}]{MillsNEB_brief}%
  \BibitemOpen
  \bibfield  {author} {\bibinfo {author} {\bibfnamefont {G.}~\bibnamefont
  {Mills}}\ and\ \bibinfo {author} {\bibfnamefont {H.}~\bibnamefont
  {J\'{o}nsson}},\ }\href {\doibase 10.1103/PhysRevLett.72.1124} {\bibfield
  {journal} {\bibinfo  {journal} {Phys. Rev. Lett.}\ }\textbf {\bibinfo
  {volume} {72}},\ \bibinfo {pages} {1124} (\bibinfo {year}
  {1994})}\BibitemShut {NoStop}%
\bibitem [{\citenamefont {Henkelman}, \citenamefont {Uberuaga},\ and\
  \citenamefont {J\'{o}nsson}(2000)}]{HenkelmanNEB}%
  \BibitemOpen
  \bibfield  {author} {\bibinfo {author} {\bibfnamefont {G.}~\bibnamefont
  {Henkelman}}, \bibinfo {author} {\bibfnamefont {B.~P.}\ \bibnamefont
  {Uberuaga}}, \ and\ \bibinfo {author} {\bibfnamefont {H.}~\bibnamefont
  {J\'{o}nsson}},\ }\href {\doibase 10.1063/1.1329672} {\bibfield  {journal}
  {\bibinfo  {journal} {J. Chem. Phys.}\ }\textbf {\bibinfo {volume} {113}},\
  \bibinfo {pages} {9901} (\bibinfo {year} {2000})}\BibitemShut {NoStop}%
\bibitem [{\citenamefont {Koistinen}\ \emph {et~al.}(2017)\citenamefont
  {Koistinen}, \citenamefont {Dagbjartsd\'{o}ttir}, \citenamefont
  {\'{A}sgeirsson}, \citenamefont {Vehtari},\ and\ \citenamefont
  {J\'{o}nsson}}]{GPRNEB_Jonsson}%
  \BibitemOpen
  \bibfield  {author} {\bibinfo {author} {\bibfnamefont {O.-P.}\ \bibnamefont
  {Koistinen}}, \bibinfo {author} {\bibfnamefont {F.~B.}\ \bibnamefont
  {Dagbjartsd\'{o}ttir}}, \bibinfo {author} {\bibfnamefont {V.}~\bibnamefont
  {\'{A}sgeirsson}}, \bibinfo {author} {\bibfnamefont {A.}~\bibnamefont
  {Vehtari}}, \ and\ \bibinfo {author} {\bibfnamefont {H.}~\bibnamefont
  {J\'{o}nsson}},\ }\href {\doibase 10.1063/1.4986787} {\bibfield  {journal}
  {\bibinfo  {journal} {J. Chem. Phys.}\ }\textbf {\bibinfo {volume} {147}},\
  \bibinfo {pages} {152720} (\bibinfo {year} {2017})}\BibitemShut {NoStop}%
\bibitem [{\citenamefont {Mills}\ and\ \citenamefont
  {Popelier}(2011)}]{IntramolMultipolKriging}%
  \BibitemOpen
  \bibfield  {author} {\bibinfo {author} {\bibfnamefont {M.~J.}\ \bibnamefont
  {Mills}}\ and\ \bibinfo {author} {\bibfnamefont {P.~L.}\ \bibnamefont
  {Popelier}},\ }\href {\doibase https://doi.org/10.1016/j.comptc.2011.04.004}
  {\bibfield  {journal} {\bibinfo  {journal} {Comput. Theor. Chem.}\ }\textbf
  {\bibinfo {volume} {975}},\ \bibinfo {pages} {42 } (\bibinfo {year}
  {2011})}\BibitemShut {NoStop}%
\bibitem [{\citenamefont {Handley}\ \emph {et~al.}(2009)\citenamefont
  {Handley}, \citenamefont {Hawe}, \citenamefont {Kell},\ and\ \citenamefont
  {Popelier}}]{polWaterKriging}%
  \BibitemOpen
  \bibfield  {author} {\bibinfo {author} {\bibfnamefont {C.~M.}\ \bibnamefont
  {Handley}}, \bibinfo {author} {\bibfnamefont {G.~I.}\ \bibnamefont {Hawe}},
  \bibinfo {author} {\bibfnamefont {D.~B.}\ \bibnamefont {Kell}}, \ and\
  \bibinfo {author} {\bibfnamefont {P.~L.~A.}\ \bibnamefont {Popelier}},\
  }\href {\doibase 10.1039/B905748J} {\bibfield  {journal} {\bibinfo  {journal}
  {Phys. Chem. Chem. Phys.}\ }\textbf {\bibinfo {volume} {11}},\ \bibinfo
  {pages} {6365} (\bibinfo {year} {2009})}\BibitemShut {NoStop}%
\bibitem [{\citenamefont {Fletcher}, \citenamefont {Kandathil},\ and\
  \citenamefont {Popelier}(2014)}]{PredKinEnergyOfCoordsKriging}%
  \BibitemOpen
  \bibfield  {author} {\bibinfo {author} {\bibfnamefont {T.~L.}\ \bibnamefont
  {Fletcher}}, \bibinfo {author} {\bibfnamefont {S.~M.}\ \bibnamefont
  {Kandathil}}, \ and\ \bibinfo {author} {\bibfnamefont {P.~L.~A.}\
  \bibnamefont {Popelier}},\ }\href@noop {} {\bibfield  {journal} {\bibinfo
  {journal} {Theor. Chem. Acc.}\ }\textbf {\bibinfo {volume} {133}},\ \bibinfo
  {pages} {1499} (\bibinfo {year} {2014})}\BibitemShut {NoStop}%
\bibitem [{\citenamefont {Ramakrishnan}\ and\ \citenamefont {von
  Lilienfeld}(2015)}]{Ramakrishnan2015}%
  \BibitemOpen
  \bibfield  {author} {\bibinfo {author} {\bibfnamefont {R.}~\bibnamefont
  {Ramakrishnan}}\ and\ \bibinfo {author} {\bibfnamefont {O.~A.}\ \bibnamefont
  {von Lilienfeld}},\ }\href@noop {} {\bibfield  {journal} {\bibinfo  {journal}
  {CHIMIA}\ }\textbf {\bibinfo {volume} {69}},\ \bibinfo {pages} {182}
  (\bibinfo {year} {2015})}\BibitemShut {NoStop}%
\bibitem [{\citenamefont {Hansen}\ \emph {et~al.}(2015)\citenamefont {Hansen},
  \citenamefont {Biegler}, \citenamefont {Ramakrishnan}, \citenamefont
  {Pronobis}, \citenamefont {von Lilienfeld}, \citenamefont {M{\"u}ller},\ and\
  \citenamefont {Tkatchenko}}]{hansen2015interaction}%
  \BibitemOpen
  \bibfield  {author} {\bibinfo {author} {\bibfnamefont {K.}~\bibnamefont
  {Hansen}}, \bibinfo {author} {\bibfnamefont {F.}~\bibnamefont {Biegler}},
  \bibinfo {author} {\bibfnamefont {R.}~\bibnamefont {Ramakrishnan}}, \bibinfo
  {author} {\bibfnamefont {W.}~\bibnamefont {Pronobis}}, \bibinfo {author}
  {\bibfnamefont {O.~A.}\ \bibnamefont {von Lilienfeld}}, \bibinfo {author}
  {\bibfnamefont {K.-R.}\ \bibnamefont {M{\"u}ller}}, \ and\ \bibinfo {author}
  {\bibfnamefont {A.}~\bibnamefont {Tkatchenko}},\ }\href {\doibase
  10.1021/acs.jpclett.5b00831} {\bibfield  {journal} {\bibinfo  {journal} {J.
  Phys. Chem. Lett.}\ }\textbf {\bibinfo {volume} {6}},\ \bibinfo {pages}
  {2326} (\bibinfo {year} {2015})}\BibitemShut {NoStop}%
\bibitem [{\citenamefont {Dral}\ \emph {et~al.}(2017)\citenamefont {Dral},
  \citenamefont {Owens}, \citenamefont {Yurchenko},\ and\ \citenamefont
  {Thiel}}]{PESandVibLevels}%
  \BibitemOpen
  \bibfield  {author} {\bibinfo {author} {\bibfnamefont {P.}~\bibnamefont
  {Dral}}, \bibinfo {author} {\bibfnamefont {A.}~\bibnamefont {Owens}},
  \bibinfo {author} {\bibfnamefont {S.}~\bibnamefont {Yurchenko}}, \ and\
  \bibinfo {author} {\bibfnamefont {W.}~\bibnamefont {Thiel}},\ }\href@noop {}
  {\bibfield  {journal} {\bibinfo  {journal} {J. Chem. Phys.}\ }\textbf
  {\bibinfo {volume} {146}},\ \bibinfo {pages} {244108} (\bibinfo {year}
  {2017})}\BibitemShut {NoStop}%
\bibitem [{\citenamefont {Li}, \citenamefont {Kermode},\ and\ \citenamefont
  {De~Vita}(2015)}]{OnTheFlyMachineLearningQMForces}%
  \BibitemOpen
  \bibfield  {author} {\bibinfo {author} {\bibfnamefont {Z.}~\bibnamefont
  {Li}}, \bibinfo {author} {\bibfnamefont {J.~R.}\ \bibnamefont {Kermode}}, \
  and\ \bibinfo {author} {\bibfnamefont {A.}~\bibnamefont {De~Vita}},\ }\href
  {\doibase 10.1103/PhysRevLett.114.096405} {\bibfield  {journal} {\bibinfo
  {journal} {Phys. Rev. Lett.}\ }\textbf {\bibinfo {volume} {114}},\ \bibinfo
  {pages} {096405} (\bibinfo {year} {2015})}\BibitemShut {NoStop}%
\bibitem [{\citenamefont {K{\"a}stner}\ \emph {et~al.}(2009)\citenamefont
  {K{\"a}stner}, \citenamefont {Carr}, \citenamefont {Keal}, \citenamefont
  {Thiel}, \citenamefont {Wander},\ and\ \citenamefont {Sherwood}}]{dlfind}%
  \BibitemOpen
  \bibfield  {author} {\bibinfo {author} {\bibfnamefont {J.}~\bibnamefont
  {K{\"a}stner}}, \bibinfo {author} {\bibfnamefont {J.~M.}\ \bibnamefont
  {Carr}}, \bibinfo {author} {\bibfnamefont {T.~W.}\ \bibnamefont {Keal}},
  \bibinfo {author} {\bibfnamefont {W.}~\bibnamefont {Thiel}}, \bibinfo
  {author} {\bibfnamefont {A.}~\bibnamefont {Wander}}, \ and\ \bibinfo {author}
  {\bibfnamefont {P.}~\bibnamefont {Sherwood}},\ }\href {\doibase
  10.1021/jp9028968} {\bibfield  {journal} {\bibinfo  {journal} {J. Phys. Chem.
  A}\ }\textbf {\bibinfo {volume} {113}},\ \bibinfo {pages} {11856} (\bibinfo
  {year} {2009})}\BibitemShut {NoStop}%
\bibitem [{\citenamefont {Sherwood}\ \emph {et~al.}(2003)\citenamefont
  {Sherwood}, \citenamefont {de~Vries}, \citenamefont {Guest}, \citenamefont
  {Schreckenbach}, \citenamefont {Catlow}, \citenamefont {French},
  \citenamefont {Sokol}, \citenamefont {Bromley}, \citenamefont {Thiel},
  \citenamefont {Turner}, \citenamefont {Billeter}, \citenamefont {Terstegen},
  \citenamefont {Thiel}, \citenamefont {Kendrick}, \citenamefont {Rogers},
  \citenamefont {Casci}, \citenamefont {Watson}, \citenamefont {King},
  \citenamefont {Karlsen}, \citenamefont {Sj{\o}voll}, \citenamefont {Fahmi},
  \citenamefont {Sch{\"a}fer},\ and\ \citenamefont {Lennartz}}]{SHERWOOD20031}%
  \BibitemOpen
  \bibfield  {author} {\bibinfo {author} {\bibfnamefont {P.}~\bibnamefont
  {Sherwood}}, \bibinfo {author} {\bibfnamefont {A.~H.}\ \bibnamefont
  {de~Vries}}, \bibinfo {author} {\bibfnamefont {M.~F.}\ \bibnamefont {Guest}},
  \bibinfo {author} {\bibfnamefont {G.}~\bibnamefont {Schreckenbach}}, \bibinfo
  {author} {\bibfnamefont {C.~A.}\ \bibnamefont {Catlow}}, \bibinfo {author}
  {\bibfnamefont {S.~A.}\ \bibnamefont {French}}, \bibinfo {author}
  {\bibfnamefont {A.~A.}\ \bibnamefont {Sokol}}, \bibinfo {author}
  {\bibfnamefont {S.~T.}\ \bibnamefont {Bromley}}, \bibinfo {author}
  {\bibfnamefont {W.}~\bibnamefont {Thiel}}, \bibinfo {author} {\bibfnamefont
  {A.~J.}\ \bibnamefont {Turner}}, \bibinfo {author} {\bibfnamefont
  {S.}~\bibnamefont {Billeter}}, \bibinfo {author} {\bibfnamefont
  {F.}~\bibnamefont {Terstegen}}, \bibinfo {author} {\bibfnamefont
  {S.}~\bibnamefont {Thiel}}, \bibinfo {author} {\bibfnamefont
  {J.}~\bibnamefont {Kendrick}}, \bibinfo {author} {\bibfnamefont {S.~C.}\
  \bibnamefont {Rogers}}, \bibinfo {author} {\bibfnamefont {J.}~\bibnamefont
  {Casci}}, \bibinfo {author} {\bibfnamefont {M.}~\bibnamefont {Watson}},
  \bibinfo {author} {\bibfnamefont {F.}~\bibnamefont {King}}, \bibinfo {author}
  {\bibfnamefont {E.}~\bibnamefont {Karlsen}}, \bibinfo {author} {\bibfnamefont
  {M.}~\bibnamefont {Sj{\o}voll}}, \bibinfo {author} {\bibfnamefont
  {A.}~\bibnamefont {Fahmi}}, \bibinfo {author} {\bibfnamefont
  {A.}~\bibnamefont {Sch{\"a}fer}}, \ and\ \bibinfo {author} {\bibfnamefont
  {C.}~\bibnamefont {Lennartz}},\ }\href {\doibase
  https://doi.org/10.1016/S0166-1280(03)00285-9} {\bibfield  {journal}
  {\bibinfo  {journal} {J. Mol. Struct. Theochem.}\ }\textbf {\bibinfo {volume}
  {632}},\ \bibinfo {pages} {1 } (\bibinfo {year} {2003})}\BibitemShut
  {NoStop}%
\bibitem [{\citenamefont {Metz}\ \emph {et~al.}(2014)\citenamefont {Metz},
  \citenamefont {K{\"a}stner}, \citenamefont {Sokol}, \citenamefont {Keal},\
  and\ \citenamefont {Sherwood}}]{ChemshellReview}%
  \BibitemOpen
  \bibfield  {author} {\bibinfo {author} {\bibfnamefont {S.}~\bibnamefont
  {Metz}}, \bibinfo {author} {\bibfnamefont {J.}~\bibnamefont {K{\"a}stner}},
  \bibinfo {author} {\bibfnamefont {A.~A.}\ \bibnamefont {Sokol}}, \bibinfo
  {author} {\bibfnamefont {T.~W.}\ \bibnamefont {Keal}}, \ and\ \bibinfo
  {author} {\bibfnamefont {P.}~\bibnamefont {Sherwood}},\ }\href {\doibase
  10.1002/wcms.1163} {\bibfield  {journal} {\bibinfo  {journal} {Wiley
  Interdiscip. Rev. Comput. Mol. Sci.}\ }\textbf {\bibinfo {volume} {4}},\
  \bibinfo {pages} {101} (\bibinfo {year} {2014})}\BibitemShut {NoStop}%
\bibitem [{\citenamefont {Mat{\'e}rn}(2013)}]{matern2013spatial}%
  \BibitemOpen
  \bibfield  {author} {\bibinfo {author} {\bibfnamefont {B.}~\bibnamefont
  {Mat{\'e}rn}},\ }\href@noop {} {\emph {\bibinfo {title} {Spatial
  variation}}},\ Vol.~\bibinfo {volume} {36}\ (\bibinfo  {publisher} {SSBM},\
  \bibinfo {year} {2013})\BibitemShut {NoStop}%
\bibitem [{\citenamefont {Rasmussen}\ and\ \citenamefont
  {Williams}(2006)}]{rasmussen2006gaussian}%
  \BibitemOpen
  \bibfield  {author} {\bibinfo {author} {\bibfnamefont {C.~E.}\ \bibnamefont
  {Rasmussen}}\ and\ \bibinfo {author} {\bibfnamefont {C.~K.}\ \bibnamefont
  {Williams}},\ }\href@noop {} {\emph {\bibinfo {title} {Gaussian processes for
  machine learning}}},\ Vol.~\bibinfo {volume} {1}\ (\bibinfo  {publisher} {MIT
  press Cambridge},\ \bibinfo {year} {2006})\BibitemShut {NoStop}%
\bibitem [{\citenamefont {Zheng}\ and\ \citenamefont
  {Frisch}(2017)}]{ZhengOptInterpolSurfaces}%
  \BibitemOpen
  \bibfield  {author} {\bibinfo {author} {\bibfnamefont {J.}~\bibnamefont
  {Zheng}}\ and\ \bibinfo {author} {\bibfnamefont {M.~J.}\ \bibnamefont
  {Frisch}},\ }\href@noop {} {\bibfield  {journal} {\bibinfo  {journal} {J.
  Chem. Theory Comput.}\ } (\bibinfo {year} {2017})}\BibitemShut {NoStop}%
\bibitem [{\citenamefont {Shepard}(1968)}]{Shepard}%
  \BibitemOpen
  \bibfield  {author} {\bibinfo {author} {\bibfnamefont {D.}~\bibnamefont
  {Shepard}},\ }in\ \href {\doibase 10.1145/800186.810616} {\emph {\bibinfo
  {booktitle} {Proceedings of the 1968 23rd ACM National Conference}}},\
  \bibinfo {series and number} {ACM '68}\ (\bibinfo  {publisher} {ACM},\
  \bibinfo {address} {New York, NY, USA},\ \bibinfo {year} {1968})\ pp.\
  \bibinfo {pages} {517--524}\BibitemShut {NoStop}%
\bibitem [{\citenamefont {Baker}\ and\ \citenamefont {Chan}(1996)}]{Baker}%
  \BibitemOpen
  \bibfield  {author} {\bibinfo {author} {\bibfnamefont {J.}~\bibnamefont
  {Baker}}\ and\ \bibinfo {author} {\bibfnamefont {F.}~\bibnamefont {Chan}},\
  }\href {\doibase
  10.1002/(SICI)1096-987X(199605)17:7<888::AID-JCC12>3.0.CO;2-7} {\bibfield
  {journal} {\bibinfo  {journal} {J. Comput. Chem.}\ }\textbf {\bibinfo
  {volume} {17}},\ \bibinfo {pages} {888} (\bibinfo {year} {1996})}\BibitemShut
  {NoStop}%
\bibitem [{\citenamefont {Dewar}\ \emph {et~al.}(1985)\citenamefont {Dewar},
  \citenamefont {Zoebisch}, \citenamefont {Healy},\ and\ \citenamefont
  {Stewart}}]{dewar1985development}%
  \BibitemOpen
  \bibfield  {author} {\bibinfo {author} {\bibfnamefont {M.~J.}\ \bibnamefont
  {Dewar}}, \bibinfo {author} {\bibfnamefont {E.~G.}\ \bibnamefont {Zoebisch}},
  \bibinfo {author} {\bibfnamefont {E.~F.}\ \bibnamefont {Healy}}, \ and\
  \bibinfo {author} {\bibfnamefont {J.~J.}\ \bibnamefont {Stewart}},\
  }\href@noop {} {\bibfield  {journal} {\bibinfo  {journal} {J. Am. Chem.
  Soc.}\ }\textbf {\bibinfo {volume} {107}},\ \bibinfo {pages} {3902} (\bibinfo
  {year} {1985})}\BibitemShut {NoStop}%
\bibitem [{\citenamefont {Sen}\ \emph {et~al.}(2015)\citenamefont {Sen},
  \citenamefont {Frey}, \citenamefont {Meisner}, \citenamefont {K{\"a}stner},\
  and\ \citenamefont {Buchmeiser}}]{Molybdenum}%
  \BibitemOpen
  \bibfield  {author} {\bibinfo {author} {\bibfnamefont {S.}~\bibnamefont
  {Sen}}, \bibinfo {author} {\bibfnamefont {W.}~\bibnamefont {Frey}}, \bibinfo
  {author} {\bibfnamefont {J.}~\bibnamefont {Meisner}}, \bibinfo {author}
  {\bibfnamefont {J.}~\bibnamefont {K{\"a}stner}}, \ and\ \bibinfo {author}
  {\bibfnamefont {M.~R.}\ \bibnamefont {Buchmeiser}},\ }\href {\doibase
  https://doi.org/10.1016/j.jorganchem.2015.09.009} {\bibfield  {journal}
  {\bibinfo  {journal} {J. Organomet. Chem.}\ }\textbf {\bibinfo {volume}
  {799-800}},\ \bibinfo {pages} {223} (\bibinfo {year} {2015})}\BibitemShut
  {NoStop}%
\bibitem [{\citenamefont {Becke}(1988)}]{bec88}%
  \BibitemOpen
  \bibfield  {author} {\bibinfo {author} {\bibfnamefont {A.}~\bibnamefont
  {Becke}},\ }\href@noop {} {\bibfield  {journal} {\bibinfo  {journal} {Phys.
  Rev. A}\ }\textbf {\bibinfo {volume} {38}},\ \bibinfo {pages} {3098}
  (\bibinfo {year} {1988})}\BibitemShut {NoStop}%
\bibitem [{\citenamefont {Perdew}(1986)}]{perdew1986}%
  \BibitemOpen
  \bibfield  {author} {\bibinfo {author} {\bibfnamefont {J.~P.}\ \bibnamefont
  {Perdew}},\ }\href {\doibase 10.1103/PhysRevB.33.8822} {\bibfield  {journal}
  {\bibinfo  {journal} {Phys. Rev. B}\ }\textbf {\bibinfo {volume} {33}},\
  \bibinfo {pages} {8822} (\bibinfo {year} {1986})}\BibitemShut {NoStop}%
\bibitem [{\citenamefont {Weigend}\ and\ \citenamefont
  {Ahlrichs}(2005)}]{wei05}%
  \BibitemOpen
  \bibfield  {author} {\bibinfo {author} {\bibfnamefont {F.}~\bibnamefont
  {Weigend}}\ and\ \bibinfo {author} {\bibfnamefont {R.}~\bibnamefont
  {Ahlrichs}},\ }\href {\doibase 10.1039/B508541A} {\bibfield  {journal}
  {\bibinfo  {journal} {Phys. Chem. Chem. Phys.}\ }\textbf {\bibinfo {volume}
  {7}},\ \bibinfo {pages} {3297} (\bibinfo {year} {2005})}\BibitemShut
  {NoStop}%
\end{thebibliography}%
	

\end{document}